\documentclass{aa} 
\pdfoutput=1 
\usepackage{graphicx}
\usepackage{amsmath}
\usepackage{amsfonts}
\usepackage[linkcolor=blue,urlcolor=magenta]{hyperref}
\usepackage[varg]{txfonts}
\bibpunct{(}{)}{;}{a}{}{,} % to follow the A&A style

\usepackage{color}

\begin{document} 

        \title{MUSE-inspired view of the quasar Q2059-360, its Lyman $\alpha$ blob,
        and its neighborhood\thanks{Based on observations collected at the European
        Organisation for Astronomical Research in the Southern Hemisphere under
        ESO programme 60.A-9331(A)}}

   \author{P. L. North\inst{1}
          \and 
          R. A. Marino\inst{2}
          \and
          C. Gorgoni\inst{1}
          \and
          M. Hayes\inst{3}
          \and
          D. Sluse\inst{4}
          \and
          D. Chelouche\inst{5}
          \and
          A. Verhamme\inst{6}
          \and
          S. Cantalupo\inst{2}
          \and
          F. Courbin\inst{1}
          }
          
\institute{Institute of Physics, Laboratory of Astrophysics, École Polytechnique Fédérale de Lausanne (EPFL),\\ 
               Observatoire de Sauverny, 1290 Versoix, Switzerland
        \and
        Institute for Astronomy, ETH Z\"urich, Wolfgang-Pauli Strasse 27, 8093 Z\"urich, Switzerland
        \and
        Department of Astronomy, Oskar Klein Centre,
        Stockholm University, AlbaNova University Centre, SE--106 91 Stockholm, Sweden
        \and
        STAR Institute, Quartier Agora -- All\'ee du six Ao\^ut 19c, B-4000 Li\`ege, Belgium
        \and
        Department of Physics, University of Haifa, Mount Carmel, Haifa 31905, Israel
        \and
        Observatoire de Gen\`eve, Universit\'e de Gen\`eve, Ch. des Maillettes 51, 1290 Versoix,
        Switzerland
        }

\date{Received  / Accepted }
 
\abstract
  {The radio-quiet quasar Q2059-360 at redshift $z=3.08$ is known to be close to a small
  Lyman $\alpha$ blob (LAB) and to be absorbed by a proximate damped Ly$\alpha$ (PDLA) system.
  
  Here, we present the Multi Unit Spectroscopic Explorer (MUSE) integral field spectroscopy
  follow-up of this quasi-stellar object (QSO). Our primary goal
  is to characterize this LAB in detail by mapping it both spatially and spectrally
  using the Ly$\alpha$ line, and by looking for high-ionization lines to constrain
  the emission mechanism.
  
  Combining the high sensitivity of the MUSE integral field spectrograph mounted on the Yepun
  telescope at ESO-VLT
  with the natural coronograph provided by the PDLA, we map the LAB down to the QSO position,
  after robust subtraction of QSO light in the spectral domain.
  
  In addition to confirming earlier results for the small bright
  component of the LAB, we unveil a faint filamentary emission protruding to
  the south over about $80$\,pkpc (physical kpc); this results in a total size of about $120$\,pkpc.
  We derive the
  velocity field of the LAB (assuming no transfer effects) and map the Ly$\alpha$ line width.
  Upper limits are set to the flux of the N\,\textsc{v}\,$\lambda 1238-1242$,
  C\,\textsc{iv}\,$\lambda 1548-1551$, He\,\textsc{ii}\,$\lambda 1640$, and
  C\,\textsc{iii}]\,$\lambda 1548-1551$ lines. We have discovered two probable Ly$\alpha$ emitters
at the same redshift as the LAB and at projected distances of $265$\,kpc and $207$\,kpc
  from the QSO; their Ly$\alpha$ luminosities
  might well be enhanced by the QSO radiation. We also find an emission line galaxy at $z=0.33$ near the line of sight
to the QSO.
  
  This LAB shares the same general characteristics as the $17$ others surrounding radio- quiet QSOs presented previously. However, there are indications that it
  may be centered on the PDLA galaxy rather than on the QSO.
  }


   \maketitle
   
\section{Introduction}

Lyman $\alpha$ (Ly$\alpha$) nebulae have been discovered
around many quasi-stellar objects (QSOs) and radio-galaxies at high redshift through slit spectroscopy
\citep{H91a,LB98,BSS03,VM03,WMF05,vB06,WCB11,NCE12}, narrow-band imaging
\citep{HC87,SAS00,SJ07,YZJ14}, and more recently, through integral field spectroscopy
(hereafter IFS) \citep{CJW06,FMD06,HWR15,BCL16}. They easily span tens of physical kiloparsecs
(pkpc) and may be larger than
$100$\,pkpc. Those that host a radio galaxy or a radio-loud QSO are brighter than
those hosting a radio-quiet QSO (RQQ), presumably because in the former, interactions between
the relativistic jet and the circum-galactic gas enhances ionization \citep{VM03}.
The Ly$\alpha$ nebulae surrounding an RQQ or that host no remarkable
object are also called Ly$\alpha$ blobs (LABs). Although they do not necessarily host a
QSO or active galactic nucleus (AGN), they are found preferentially in overdense regions
\citep{EBS11,MYHT04,MNM09,MYH11,PKD08,YZTE09}.
These nebulae are interesting because they trace the
circum-galactic medium (CGM) at high redshifts, thereby probing galaxy formation, especially
if their Ly$\alpha$ fluorescence is enhanced by the ultraviolet (UV) radiation of the QSO, as
outlined by \citet{HR01} and \citet{CPL05,CLP07,CLH12}.

Although most LABs remain confined within the virial radii of their respective haloes,
recent observations have uncovered remarkably large nebulae extending beyond the virial
radius \citep{CAP14,MCM14a,MCM14b,MMM15,HPC15}, with typical sizes of $400$~pkpc. These
outstanding observations, some of them made possible by new-generation IFSs
like the Palomar Cosmic Web Imager \citep{MCM14a,MCM14b,MMM15}, provide the first evidence
of the cosmic web filaments predicted by large-scale $\Lambda$CDM simulations.

The frequency of LABs around RQQs has remained an open question \citep[e.g.,][]{NCE12} until
the new-generation IFS Multi Unit Spectroscopic Explorer (MUSE) attached to the $\text{}$fourth unit of the VLT
at ESO was used to observe 17 RQQs at redshifts $3 < z < 4$ and found an LAB around each
of them \citep[][hereafter B16]{BCL16}. Therefore, in this redshift range and for a sensitivity of about
$10^{-18}$\,erg\,s$^{-1}$\,cm$^{-2}$\,arcsec$^{-2}$, the frequency of LABs surrounding
RQQs is $100$\%. Moreover, the size of such nebulae always reaches or exceeds $100$~pkpc,
and their surface brightness (SB) profiles are consistent with a power law with
a slope of about $-1.8$.

Obviously, the ubiquity and similarity of the LABs that enshroud RQQs raise the question
of their power source and emission mechanism. In addition to fluorescent
emission from photo-excitation or photoionization \citep[e.g.,][]{GALS09},
other power sources have been suggested, like cooling radiation
where gravitational energy of the gas collapsing in the potential well of the dark matter halo
is converted into Ly$\alpha$ radiation after collisional excitation and ionization
\citep[known also as the ``cold accretion'' scenario, see, e.g.,][]{HSQ00,FKG01,DL09,RB12}.
Shock heating was also proposed \citep{TS00,MUF04}. Finally, scattering of the Ly$\alpha$
photons emitted by a central source (like an AGN) may also occur, betrayed by a high
level of linear polarization in the external parts of the nebula \citep{LA98,RL99,DL08,DL09,DK12}.
This effect was first observed by \citet{HSS11} and may coexist with the cold accretion
scenario \citep{TVB16}. Additional lines such as \ion{He}{ii}$\lambda 1640$ can
potentially help distinguish between the cold accretion and the photoionization
scenarios, although very deep obervations seem to be needed \citep{ABY15}.

Neutral hydrogen clouds may lie on the line of sight of a QSO and produce a so-called
damped Lyman alpha (DLA) absorption system, as soon as the column number density exceeds
$2\times 10^{20}$\,cm$^{-2}$. When the velocity difference between the DLA and the
background QSO is $\Delta v < 3000$\,km\,s$^{-1}$, the two objects are considered
as being ``associated'' and the DLA is then classified as a proximate damped Lyman alpha
(PDLA) system \citep{EYH02,EPH10}. Even though the association may not be physical in the
gravitational sense, it can be manifested through enhanced incidence of highly
ionized species, for example, showing that the absorber is ``influenced by the radiation field of
the QSO'', as \citet{R01} phrased it.
The frequency of PDLAs is close to $2$\% of the QSOs observed in the redshift range
$2.2<z<5$, and at redshift $3$, it exceeds the frequency expected assuming that they
are drawn from the same population as the intervening DLAs \citep{PHH08,EPH10}.
In addition to its intrinsic interest, a PDLA provides
the advantage of acting like a natural coronagraph that extincts the QSO light in 
the vicinity of the Ly$\alpha$ line, allowing us to detect and study any emission
nebula surrounding the QSO more easily \citep{HPKZ09}. This configuration, where the PDLA lies
in front of the QSO and its Ly$\alpha$ nebula, in principle offers the possibility
to ``image the PDLA {\sl in silhouette} against the extended screen of Ly$\alpha$ emission'',
as outlined by \citet{HPKZ09}.

In this work we focus on an LAB lying along the line of sight of the QSO
\object{Q2059-360} (RA 21$^\textnormal{h}$ 02$^\textnormal{m}$ 44$^\textnormal{s}.7$,
Dec -35$^\circ$ 53$'$ 06.5$''$, V$=18.6$) at redshift $z = 3.09$ \citep{WHO91}\footnote{This is the
original reference. The redshift value is mentioned as uncertain for this particular
object, while the average error of most other objects is announced to be better than $0.01.$}.
The LAB is emitting at $\sim 4970$\,\AA, within the damped Ly$\alpha$ absorption (DLA)
trough of the QSO spectrum. 
The DLA that we analyze in this work is at redshift $z_{abs} = 3.082$,
that is, only $\sim 500$\,km\,s$^{-1}$ from the background QSO, if the latter is indeed
at redshift $z = 3.09$; our revised redshift estimate is rather $z = 3.092-3.095$
(see Sects. 3.1 and 3.3 below), implying a velocity difference of $700-930$\,km\,s$^{-1}$
(see Sect. 3.2 below).
This would represent
$2.3-3.0$ proper Mpc if the velocity difference were entirely due to the Hubble
flow\footnote{\url{http://home.fnal.gov/~gnedin/cc/}}. An early study of this QSO
and of its Ly$\alpha$ features has been carried out by \citet{LR99}. Based on
long-slit spectra, they showed that the emission feature is
extended both spatially and spectrally, and they provided evidence for a shift of the central
wavelength as a function of the offset position from the QSO. Here, we take advantage
of the IFS to provide a detailed 2D mapping of the structure.

In our analysis, we assume a $\Lambda$CDM universe model with parameters
$H_0=69.6$\,km\,s$^{-1}$Mpc$^{-1}$, $\Omega_M=0.286$, and $\Omega_\Lambda=0.714$ \citep{BLW14},
and we use the cosmology calculator made available by \cite{W06} on the World Wide
Web\footnote{\url{http://www.astro.ucla.edu/~wright/CosmoCalc.html}}.

\section{Observations and data reduction}
The MUSE instrument is an IFS
located at the Nasmyth B focus of Yepun, the VLT UT4 telescope. It is operating in the
visible wavelength range from 4800\,\AA$ $ to 9400\,\AA. Thanks to its wide field of view,
$1\arcmin\times 1\arcmin$ in the wide field mode, it is able to  take $90\,000$
simultaneous spectra with a spatial sampling of $0.2$\arcsec per pixel (also referred
to as "spaxel").
MUSE has a resolving power $R = 1835$ \citep{RB14} at the redshifted Ly$\alpha$ wavelength,
which in terms of velocity separation corresponds to a resolution $\Delta v$ of
\[\Delta v = c/R \approx 163 \textnormal{ km\,s$^{-1}$}\]
where $c$ is the speed of light in the air. Or equivalently in terms of wavelength,
\[\Delta\lambda = \lambda/R \approx 2.7$\,\AA,$\]
where $\lambda \sim 4970$\,\AA$ $ is the wavelength of the Ly$\alpha$ blob emission
at redshift $3.08$.

The observations were made in June 2014 during the science verification phase
(ESO program 60.A-9331(A)).
The total exposure time amounts to 3 hours 20 minutes, distributed in four observing blocks,
each of which included two $1497$\,s exposures taken at instrumental position angles differing
by $90^\circ$. Dithering of a few arcseconds was also applied.

The data set was reduced using both the ESO MUSE pipeline (version 1.6, Weilbacher 2015)
and the CubExtractor package (CubEx hereafter, version 1.6; Cantalupo, in preparation).
In particular, the basic reduction steps of bias and flat-fielding corrections, wavelength
solution, illumination correction, and flux calibration  were
made for each of the individual exposures with the ESO pipeline using the default parameters. After the calibration files
were processed, the raw science data were then reduced using the {\it scibasic} and {\it scipost}
recipes in order to obtain the datacubes and pixel tables for each exposure sampled in a
common grid of $0.2\arcsec\times 0.2\arcsec\times 1.25$\,\AA. Then, we made use of the
CubEx package to compute a more accurate astrometry, to improve the quality of the
flat-field correction (CubeFix tool), and to perform a flux-conserving sky subtraction
on the individual flatfield-corrected cubes (CubeSharp). We refer to B16
for a detailed description of the CubEx package. Finally, the eight individual exposures were combined
into a final cube using a $3\sigma$ clipping and mean statistics. 
The resulting point-spread function (PSF) measured at $7000$\,\AA\ on the final combined
cube has a full-width at half-maximum (FWHM) of $1.03$\arcsec. At the wavelength of the emitter
($4968$\,\AA), it is $1.06$\arcsec.

\begin{figure}
 \resizebox{\hsize}{!}{\includegraphics{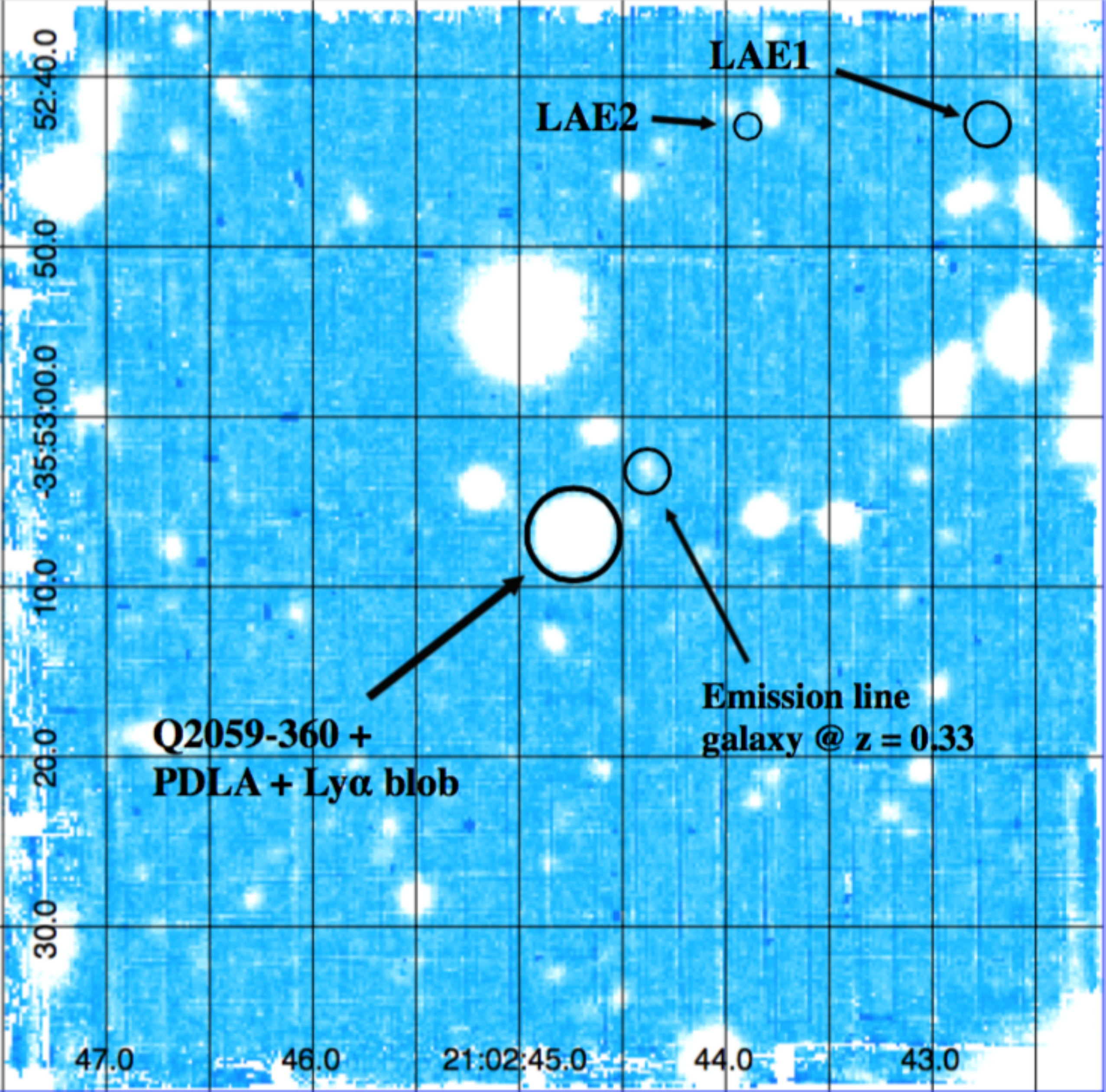}}
 \caption{White-light image obtained by integrating the datacube from $4980$\,\AA\ to $9350$\,\AA.
 The positions of the four objects discussed in this paper are indicated. The field
 of view is one arcminute on a side.}
 \label{fig:field}
\end{figure}

The white-light image obtained by collapsing the datacube along the wavelength axis
from $4980$\,\AA\ to $9350$\,\AA\  is shown in Fig.~\ref{fig:field}. The position of
the QSO is indicated, as well as that of the LAEs found at same redshift, and of an emission
line galaxy at $z=0.33$. We examine these four objects in turn below.

\section{Analysis of the datacube in the QSO vicinity}
First of all, we have redetermined the redshift of the QSO (Sect. \ref{sect:zQSO}),
and the redhift of the PDLA using metallic absorption lines (Sect. \ref{sect:zPDLA}).
Then, we have performed two independent analyses of the immediate neighborhood of the QSO, aimed
first at characterizing the LAB. The analysis is complicated by the velocity offset between
the PDLA absorption trough and the nebular emission, causing the latter to fall not in
the middle of the former, but on its red wing. Thus, the QSO contribution has to be subtracted,
even though it remains relatively small, at least in its immediate vicinity, namely within
$1$\arcsec$-2$\arcsec\ of it. We used the method proposed by \citet{LR99},
which consists of defining the QSO intrinsic spectrum and subtract it in the wavelength
domain for each spaxel; for the sake of consistency, we also performed an analysis similar
to that presented by B16 (subtraction of the QSO PSF in the
spatial plane for each relevant wavelength bin). We describe these two methods in more
detail below (Sects. \ref{sect:fitLya} and \ref{sect:subPSF}).

\subsection{\label{sect:zQSO}Redshift of the QSO}
\begin{figure*}
 \resizebox{\hsize}{!}{\includegraphics[width=\textwidth]{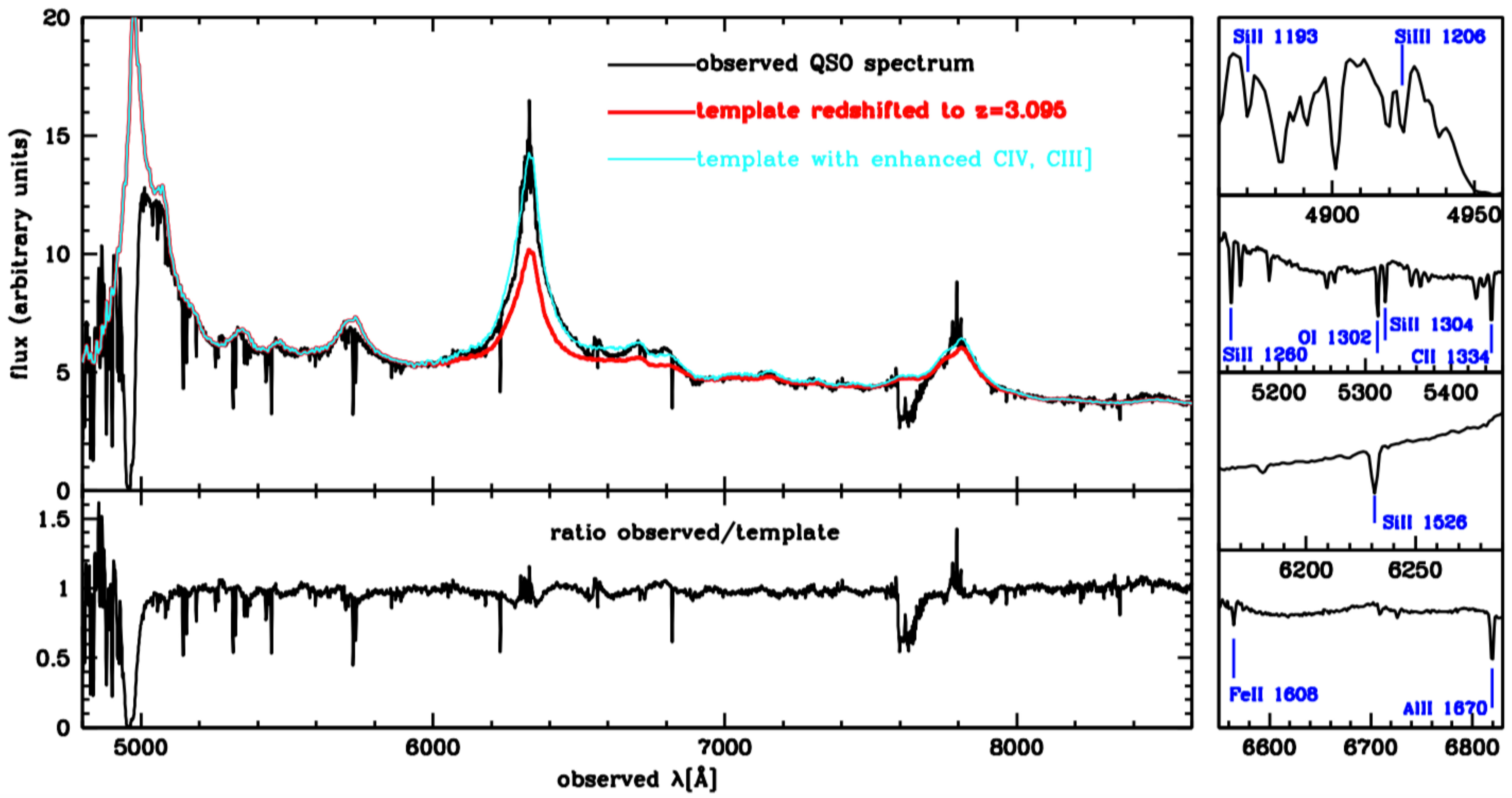}}
 \caption{{\sl Top left:} Observed QSO spectrum (black curve) compared to the normalized SDSS
 template of Vanden Berk et al. (2001) redshifted to $z=3.095$ (red curve), and to the
 same template after enhancement of the carbon emission lines (cyan curve, see text).
 {\sl Bottom left:} Ratio of the observed QSO spectrum to the redshifted and normalized
 template with enhanced C lines. {\sl Right panels:} Enlarged portions of the QSO spectrum
 showing the metallic lines used to determine the redshift of the PDLA.}
 \label{fig:redshift_QSO}
\end{figure*}

We compared the QSO spectrum extracted from the MUSE datacube with the template QSO
spectrum built by \citet{VBRB01} on the basis of SDSS spectra. The latter have about the
same resolution as MUSE, making them appropriate for the task. The most useful emission line
for redshift determination is \ion{C}{iv}$\lambda 1549$, although other features such as
\ion{Si}{iv}$\lambda 1396$/\ion{O}{iv]}$\lambda 1402$ and \ion{C}{iii]}$\lambda 1908$/\ion{Fe}{iii}
can also help. After enhancing the line amplitude with respect to the continuum in the
template spectrum (by a factor of $1.8$ for \ion{C}{iv} and 1.2 for \ion{C}{iii]}),
the best visual match is obtained for
\begin{equation}
z\approx 3.095\pm 0.003,
\end{equation}
as shown in Fig.~\ref{fig:redshift_QSO}
This is slightly larger than $z=3.09$ estimated by \citet{WHO91}, but within their $0.01$
error bar. Our redshift estimate is largely based on the \ion{C}{iv} line, which is blueshifted
on average by $810$\,km\,s$^{-1}$ with respect to the systemic velocity for radio-quiet
quasars \citep{RKG11}; however,
this should also be the case of the template spectrum, so this is not likely to cause
any severe bias.

\subsection{Redshift of the PDLA according to metallic absorption lines}\label{sect:zPDLA}
\begin{figure}
 \resizebox{\hsize}{!}{\includegraphics{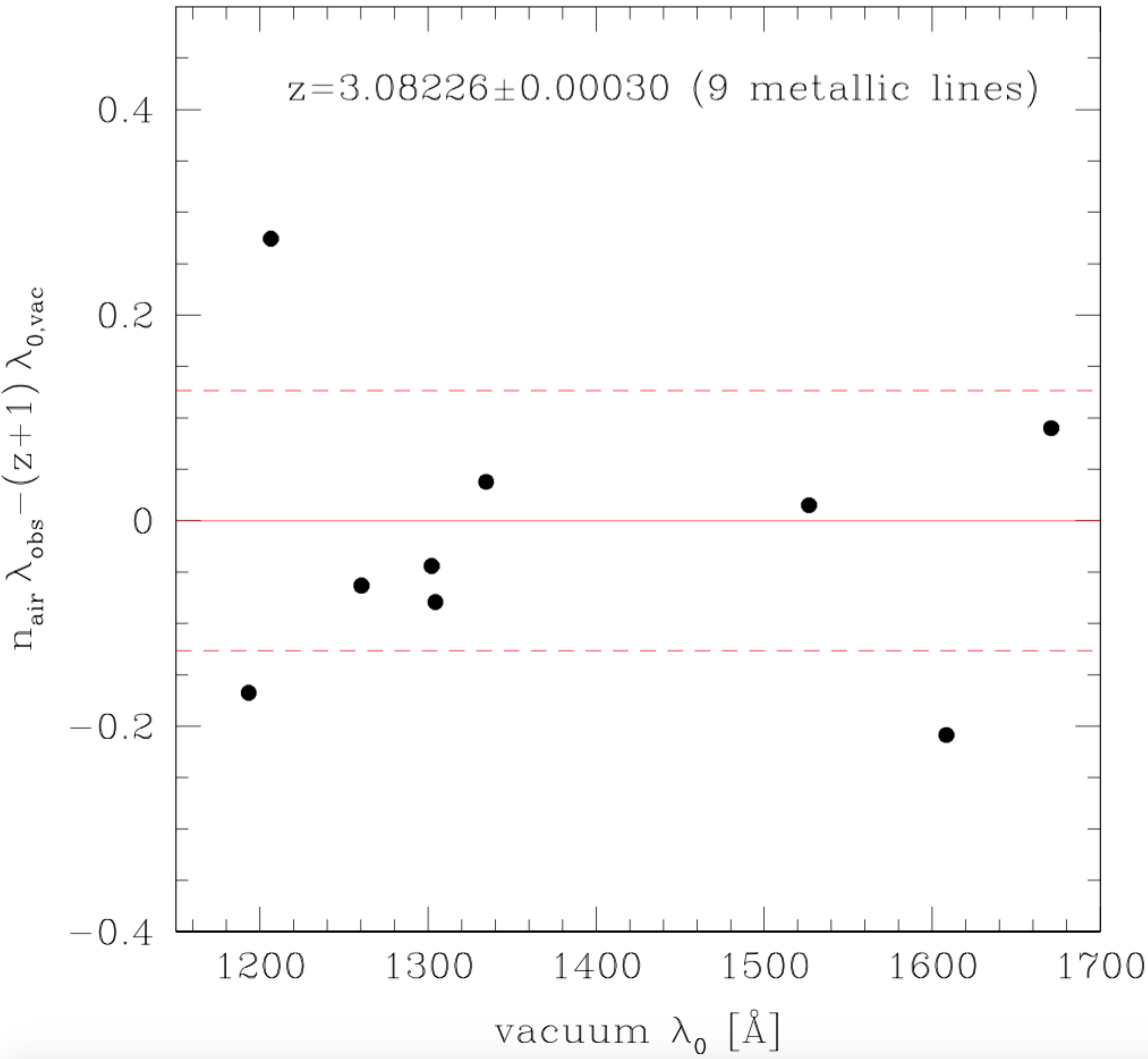}}
 \caption{Difference between the observed wavelength corrected for the air refraction
 index and the redshifted laboratory (vacuum) wavelength versus laboratory wavelength
 for the nine lines mentioned in the text. The red horizontal solid line corresponds to the
 EW-weighted regression, while the red dashed lines show the $\pm 1\sigma$ rms scatter.}
 \label{fig:zmetal}
\end{figure}

Ten metallic absorption lines can be easily identified in the PDLA system within the MUSE
spectral range. These are \ion{Si}{ii}\,$1193$, \ion{Si}{iii}\,$1206$,
\ion{N}{v}\,$1238$, \ion{Si}{ii}\,$1260$, \ion{O}{i}\,$1302$,
\ion{Si}{ii}\,$1304$, \ion{C}{ii}\,$1334$, \ion{Si}{ii}\,$1526$,
\ion{Fe}{ii}\,$1608$, and \ion{Al}{ii}\,$1670$. The \ion{N}{v}\,$1238$
line is the faintest of all and has the largest residual from the regression of observed
versus laboratory wavelengths, so we discarded it. We also discarded the
\ion{C}{iv}\,$1548-1550$ lines because they fall in the middle of telluric lines.
The difference between the observed
wavelength corrected for the air refraction index ($n_\mathrm{air}\sim 1.00205$)
and the redshifted vacuum wavelength is shown in Fig.~\ref{fig:zmetal}. A least-squares fit
weighted by the observed equivalent widths (EW) of the $\text{nine}$ lines gives a redshift
\begin{equation}
z_\mathrm{abs}=3.08226\pm 0.00030
\end{equation}
with an rms scatter of  $0.127$\,\AA, much smaller than the spectral resolution. Our
estimate agrees with the value of $3.08303$ given by \citet{LR99} (based on the
\ion{Fe}{ii}\,$1144$, \ion{Si}{ii}\,$1193$, and \ion{Si}{iii}\,$1206$ lines) to within
$1.5\,\sigma$.

There is no sign of any multiple absorption components.

The redshift difference of the PDLA relative to the QSO can be translated into a velocity
difference using the formulae given by \citet{DS14}: An object with a ``peculiar'' redshift
$z_p$ with respect to a QSO at redshift $z_{em}$ will appear to the observer to have $z$ such that
\begin{equation}
1 + z = (1 + z_{em})(1 + z_p) \implies z_p = \cfrac{z-z_{em}}{1+z_{em}}\approx -0.00311
\label{eq1}
.\end{equation}
The peculiar velocity $v_p$ is linked to $z_p$ in the following way for relativistic and non-relativistic velocities, respectively:
\begin{equation} \label{eq2}
v_p = c\,\cfrac{(1+z_p)^2 - 1}{(1+z_p)^2 + 1} \simeq c\,z_p=-930\pm 220~\mathrm{km\,s}^{-1}
,\end{equation}
where the error reflects the uncertainty on the QSO redshift.
Thus we have $|v_p| < 3000$\,km\,s$^{-1}$, justifying the classification of the absorption
system as a PDLA.

\subsection{Fit of the Lyman alpha absorption trough and subtraction of the QSO contribution}
\label{sect:fitLya}
From the whole datacube given by the pipeline, we wish to extract the spectrum of the LAB alone
in the vicinity of the QSO. The goal is then to subtract the absorption
profile from the observed spectrum to obtain a clear signal of the emission feature
without the contribution of the QSO. We first used the \texttt{ds9} tool to determine the
position of the LAB and roughly estimate its extent. This allowed us to select a region
encompassing most of the LAB luminosity, and of course the whole QSO contribution. The region
we selected is elliptical, with semi-axes of $2.3$\arcsec\ and $2.8$\arcsec, the position
angle (orientation with respect to the north direction, counted positively to the east)
of the semi-major axis being $30$\degr.
The procedure to isolate the spectrum of the LAB is then the following:

$\bullet$ We use the software \texttt{VPFit} \citep{CW14} to fit a Voigt profile to the
absorption trough. The needed error vector was extracted from the variance datacube, and we
dropped all irrelevant data, namely ($i$) absorption lines of the Ly$\alpha$ forest, on the
blue side of the DLA trough, ($ii$) some faint absorption features to the red side of the
DLA trough ($5000\,\AA<\lambda<5100$\,\AA), ($iii$) some narrow absorption lines on the blue wing
of the DLA trough ($\sim 4930\,\AA<\lambda<4940$\,\AA), and ($iv$) the Ly$\alpha$ blob emission.
These regions are shown in Fig.~\ref{fig:fit}.

In order to obtain a meaningful fit, we need to reconstruct the original energy distribution of the
QSO (we call it ``the continuum'' somewhat abusively, because here lies the broad Ly$\alpha$
line). This is done by iteration over the following steps:
\begin{figure}
 \hspace{-0.1cm}
 \resizebox{\hsize}{!}{\includegraphics{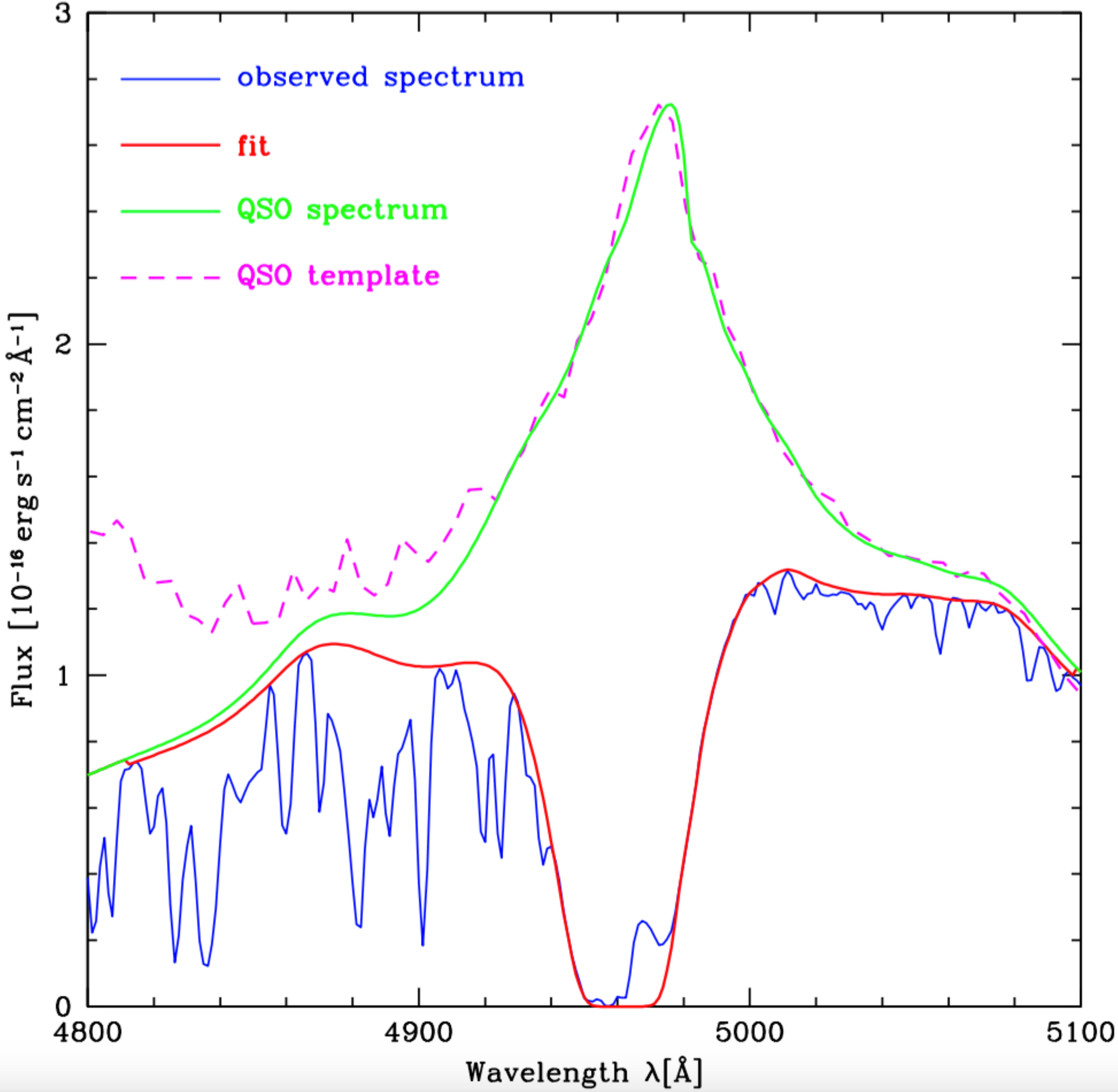}}
 \caption{QSO spectrum with the Ly$\alpha$ DLA feature. The blue line is the
 measured spectrum, with the emission feature at the bottom of the damped Ly$\alpha$
 absorption trough. The Ly$\alpha$ forest lies on the blue side of the DLA Ly$\alpha$
 absorption. The red curve shows the Voigt fit of the DLA. The green curve is the reconstructed
 "continuum", actually the broad Ly$\alpha$ emission line of the QSO (see text). Note that
 the reconstructed QSO line is only an interpolated guess between $\sim4950$~\AA\ and
 $\sim4958$~\AA. The thin dashed magenta line is the SDSS QSO template of Vanden Berk et al.
 (2001) redshifted to $z=3.092$ and normalized to match the green curve as well as possible.}
 \label{fig:fit}
\end{figure}

$\bullet$ We start by performing a first fit, feeding \texttt{VPFit} with our measured spectrum
(blue curve in Fig.~\ref{fig:fit}) and a preliminary continuum (close to the green curve
in Fig.~\ref{fig:fit}). \texttt{VPFit} will then provide a Voigt profile fit of the DLA.

\begin{figure}
 \resizebox{\hsize}{!}{\includegraphics{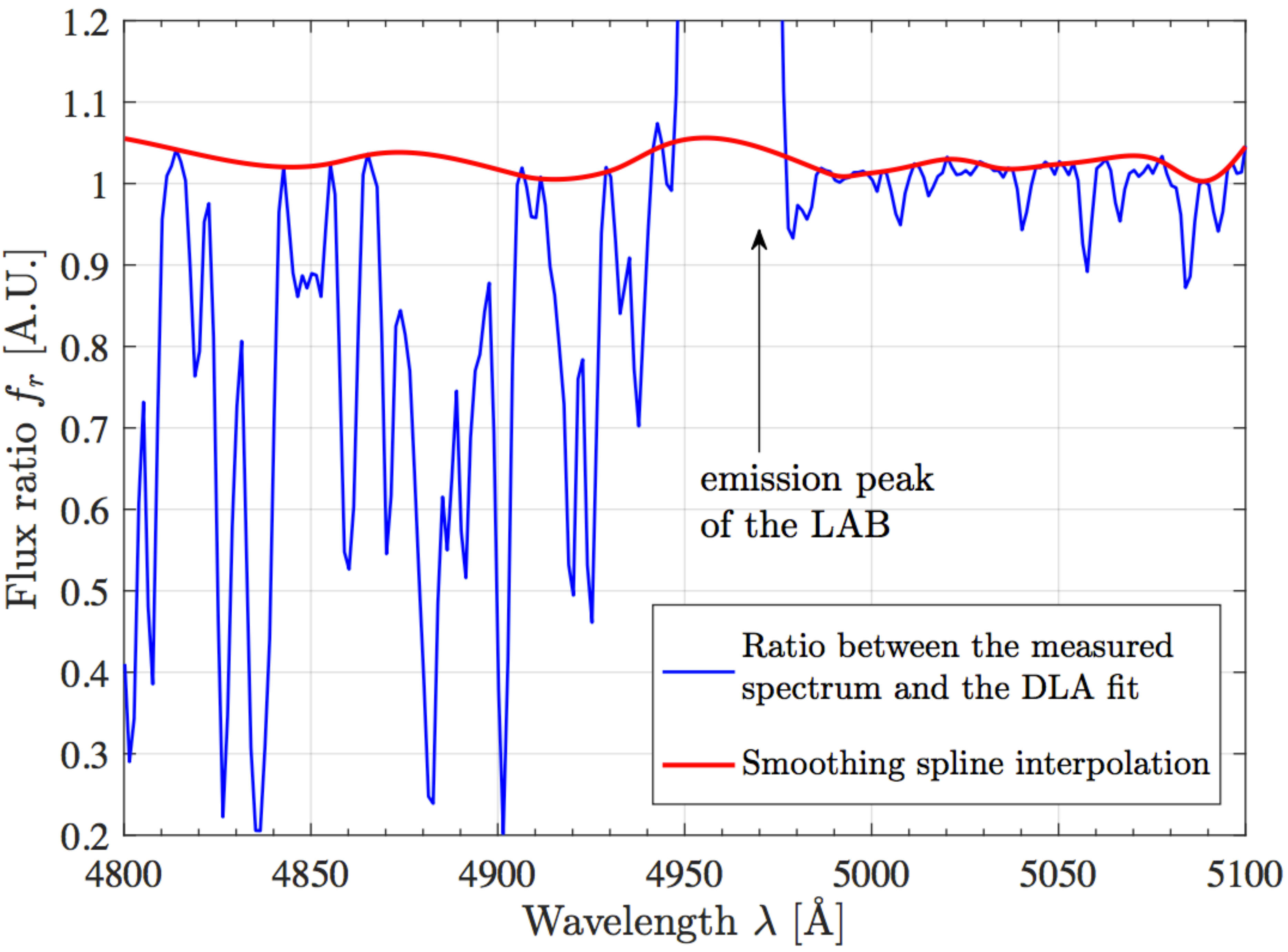}}
 \caption{Illustration of the second and third step of the procedure for estimating
 the QSO continuum. The highest values of $f_r$ (blue curve) must tend to $1$ (outside
 the bottom of the PDLA absorption and the LAB emission), when the
 profile for the DLA converges to a good solution; the blue curve diverges in the
 wavelength range corresponding to the bottom of the PDLA absorption and to the LAB
 emission, where it has to be interpolated. The red curve shows the spline
 interpolation performed over the highest points of the blue curve. The spline function
 has then to be multiplied by the previous estimate of the continuum to create a better estimate.}
 \label{fig:expl}
\end{figure}

$\bullet$ We divide the original spectrum by the DLA fit, obtaining the function $f_r$
shown in Fig.~\ref{fig:expl} (blue line). The upper envelope of $f_r$ should tend toward $1$
outside the emission region of the LAB (and outside the bottom of the absorption trough)
because one has to ensure that the wings of the Ly$\alpha$ absorption are well fitted,
regardless of the other absorption lines.
The closer $f_r$ is to $1$, the better is the continuum.

$\bullet$ We then manually define the upper envelope of
the $f_r$ function, neglecting both the bottom of the PDLA absorption (where $f_r$ is
not defined) and the LAB emission (see Fig.~\ref{fig:expl}, red line). This provides a vector
by which we multiply the previous estimate of the continuum to obtain an improved estimae.
We note that by construction, this vector remains uncertain in the wavelength range
$\sim 4948$\,\AA$-4977$\,\AA, where we are forced to interpolate the continuum. If the latter were to
present any discontinuity or peak within that range, neither this method nor any other would
be able to unveil it.

Four such iterations allowed us to converge to a satisfactory fit of the QSO+PDLA spectrum,
shown in Fig.~\ref{fig:fit}. We note that the shape of the resulting broad emission
Ly$\alpha$ line is remarkably close to that of the template spectrum, after the
latter has been properly normalized (i.e., divided by the linear function
$y=0.0035\cdot\lambda$\,[\AA]$-10.821$).

We can now scale the fitted spectrum and subtract it from the observed spectrum in each spaxel
concerned by the QSO contribution. This is done in the following way.
Let $I_{ref}$ be the spectrum extracted in the first part (the blue curve in Fig.~\ref{fig:fit}),
and $s_{ij}$ the spectrum in a particular spaxel located at position $(i,j)$ in the image.
The scaling factor $h_{ij}$ for a specific spectrum is given by the ratio between the sum of
$s_{ij}$, selected in a small wavelength range $D_\lambda=[5018.75$\,\AA$,5036.25$\,\AA] where
the spectrum is nearly flat, and the sum of $I_{ref}$ in the same range:
\begin{equation}
h_{ij} = \frac{\displaystyle\sum_{\lambda \in D_\lambda} s_{ij}(\lambda) \Delta\lambda}
               {\displaystyle\sum_{\lambda \in D_\lambda} I_{ref}(\lambda)\Delta\lambda}
\label{eq:hij}
,\end{equation}
where $\Delta \lambda = 1.25$~\AA\ is the spectral sampling of our image.
Thanks to the scaling factor, we can estimate a fit for the PDLA profile in each spaxel. If $F_{ref}$
is the QSO+PDLA spectrum computed above (red curve in Fig.~\ref{fig:fit}), the QSO+PDLA spectrum
$f_{ij}$ in each spaxel is given by
\begin{equation}
f_{ij} = h_{ij}F_{ref}
\label{eq:scaling}
.\end{equation}
Therefore, the LAB spectrum $\tilde{s}_{ij}$ in each spaxel, without the QSO contribution,
is given by subtracting the estimated fit $f_{ij}$ to the measured spectrum $s_{ij}$:
\begin{equation}
\tilde{s}_{ij} = s_{ij} - f_{ij}
\label{eq:subtract}
.\end{equation}
We now have a new subcube $\tilde{s}_{ij}$ from which the signal of the QSO has been subtracted.

\section{Exploration of the datacube at larger scales}
We have explored the datacube at scales larger than just a few arcseconds from the
QSO, both visually and in an automated way.

Visual exploration was made by scanning the cube in increasing wavelength bins and by
displaying the spectrum of the few emission
line objects that were detected. It also allowed us to discover a faint extension of the
LAB protruding to the south-southwest for some $10$\arcsec.

Automated exploration aimed at defining the extension of the LAB in a more objective and
sensitive way was performed with CubEx, which in addition to its post-processing capabilities,
is also able to exploit simultaneously the 3D information of the datacube to automatically
extract extended emission above some predefined signal to noise ratio (S/N) threshold.
More details are given hereafter and by B16. This method allowed us to deeply explore the
filamentary structure we had assumed to be there based on visual exploration, which is oriented to the south over more than $10$\arcsec\ ($\sim100$\,pkpc), and to better define
its shape and surface brightness.

\subsection{\label{sect:subPSF}PSF quasar subtraction from each monochromatic image}
We also explored an alternative empirical method that is implemented in CubEx to minimize the
contribution of the QSO
PSF and obtain the ``pure'' extended Lyman alpha emission. The principle of this method
has been widely used in imaging astronomy \citep{HWS13} and
is very well suited to IFS data \citep{HWR15}. In summary, an empirical PSF is scaled and
subtracted from the QSO at each wavelength layer. We used the CubePSFSub tool, which is also part
of the CubEx package, to create the pseudo-NB images ($150$ spectral pixels or $187$\,\AA)
at the position of the QSO, then the flux measured in the $1$\arcsec$\times 1$\arcsec\ 
central region was rescaled in each empirical PSF image and finally subtracted from the
corresponding wavelength layer in the datacube (see Sect. 3.1 in B16 for further details).
Although this method produces excellent results far away from the quasar, it does not provide
robust results in the central $1$\arcsec\ region.
Ultimately, we used the latter CubEx approach to detect the extended LAB and the
approach described in Sect. \ref{sect:fitLya} to study the PDLA properties.

\section{Results}

\subsection{Properties of the PDLA and of the bright component of the Lyman alpha blob}\label{res_lab}
We first focus on the properties of the PDLA obtained through the \verb+VPFit+ code.
Together with the fitted absorption profile, both the column density and the redshift of
the PDLA are derived:
\begin{equation}
\log(N) = 20.9290 \pm 0.0024 ~~~;~~~ z_{abs} = 3.082097 \pm 0.000053
\label{eq:col_dens_zabs}
.\end{equation}
The redshift value obtained here differs by no more than $0.000163$ from the value obtained
based on metallic lines and is thus entirely consistent with it (see Sect. 3.2 above).
We took the refractive index $n$ of air into account that causes a shift in the observed
wavelengths: $\lambda_{air} = \lambda_{vac}/n$. We used the Ciddor equation \citep{C96},
taking the atmospheric conditions at the time of the measurement
into account (air temperature
12$^\circ$C, pressure $744.9$\,hPa, and relative humidity $3$\%, as reported in the
header of the original FITS file). The errors are the formal errors provided by
\texttt{VPFit} and are certainly underestimated, as they do not include
uncertainties linked to the continuum determination.
In their pioneer study, \citet{LR99} obtained a column density and redshift
\begin{equation*}
\log(N) = 20.85 \pm 0.03 ~~~;~~~ z_{abs} = 3.0825.
\end{equation*}
Our estimate of the column density is higher than theirs by
$2.6\sigma$ (referring to their error estimate), and we find a slightly
lower redshift (the difference amounts to $7.6\sigma$ according to our own error estimate).
The latter discrepancy might be due to our lower spectral resolution ($R\sim 1840$
instead of $3300$), which might cause the metallic absorption lines in the blue wing of the
Ly$\alpha$ absorption trough to be less easily recognized and be partly included in the
Voigt profile fit. This would result in a slightly overestimated column densitiy and in a
slight blue shift. However, our \verb+VPFit+ estimate of the redshift, based on the
Ly$\alpha$ line alone, agrees within only $0.5\sigma$ with the PDLA redshift estimate based on
metallic lines.

\begin{figure}
 \hspace{-0.2cm}
 \resizebox{\hsize}{!}{\includegraphics{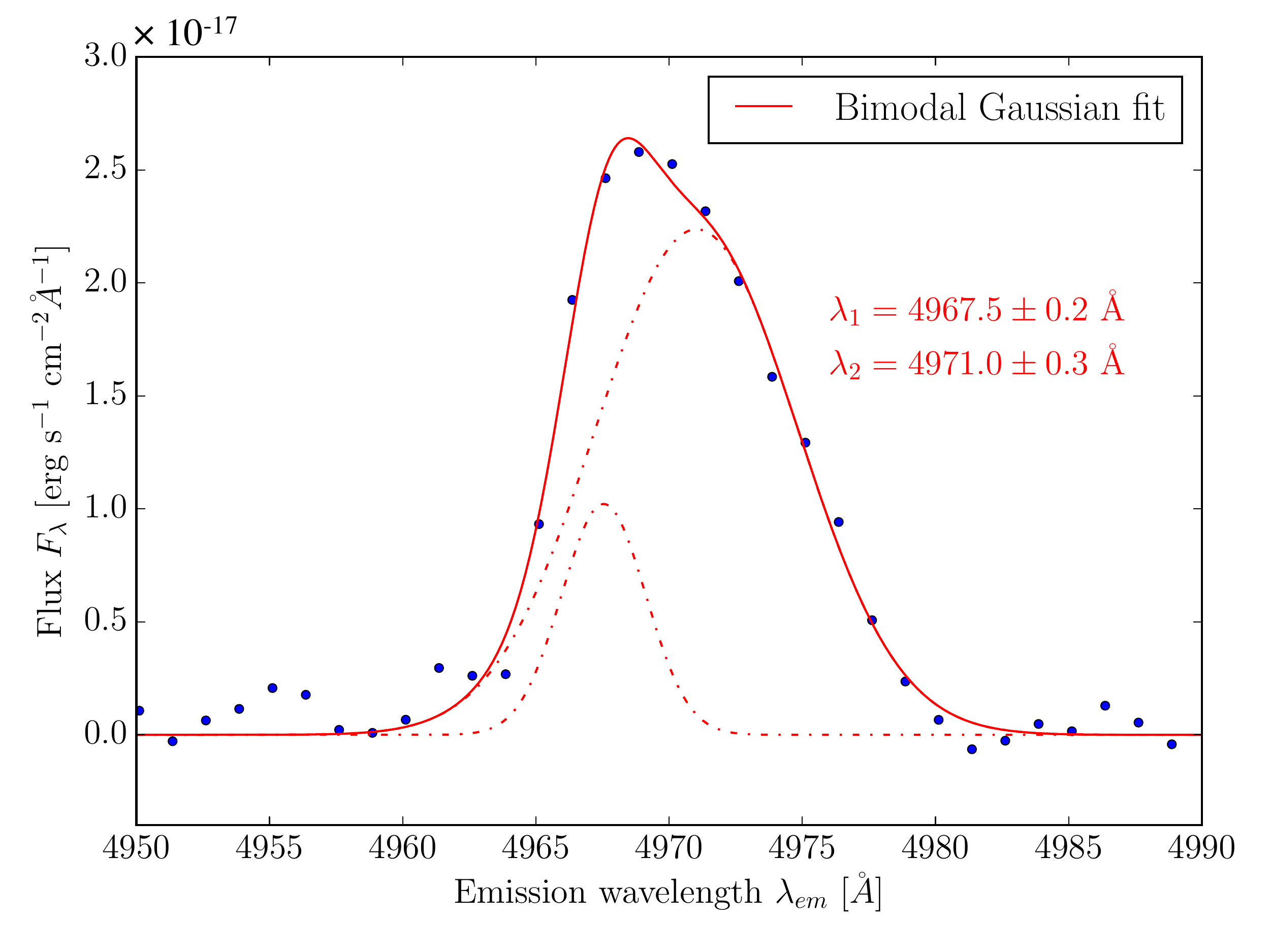}}
 \caption{Spectrum of the emission feature averaged over an elliptic surface of
 $2.3$\arcsec$\times 2.8$\arcsec, centered on the QSO, but free of its flux. The heavy continuous
 curve is a fit performed with the sum of two Gaussians (dash-dotted lines). The wavelengths
 indicated are corrected to vacuum.}
 \label{fig:lab_spec}
\end{figure}

We can determine the average spectral shape of the LAB emission\ feature over the
$4.6\arcsec\times 5.6$\arcsec\ elliptic surface defined previously by subtracting the
DLA fit to the measured spectrum. The result, shown in Fig.~\ref{fig:lab_spec}, is a clearly
asymmetric line that can formally be fitted by a pair of Gaussian functions (Fig.~\ref{fig:lab_spec}).
The parameters of the two Gaussians are listed in Table~\ref{tab_vel}. We note that these
two functions are only used as a convenient way of representing the average emission line
and do not bear straightforward physical meaning.
\begin{table}
\caption{Central position $\lambda$ (corrected to vacuum) and FWHM
of the two Gaussian components in the LAB spectrum. The third line gives the same quantities
for the total emission profile.} 
\label{tab_vel}
\centering
\begin{tabular}{lccc}
        \hline \hline
        & $\lambda$ & \small{FWHM} & \small{FWHM} \\ 
        & [\AA] & [km\,s$^{-1}$] & [\AA]\\
        \hline\\
        first peak & $4967.5 \pm 0.2$ & $225\pm 42$ & $3.7\pm 0.7$ \\
        second peak & $4971.0 \pm 0.3$ & $538\pm 24$ & $8.9\pm 0.4$ \\ \hline
      \\ first moment                 & $4970.5 \pm 0.1$ & $559\pm 12$ & $9.27\pm 0.20$ \\
        \hline
\end{tabular}
\end{table}
The total Ly$\alpha$ flux $F_{Ly\alpha}$ of the emission feature is computed by integrating
the flux over the double Gaussian profile:
\begin{equation}
\label{eq:total_flux}
F_{Ly\alpha} = (2.53 \pm 0.05)\cdot10^{-16} \textnormal{ erg s$^{-1}$ cm$^{-2}$}
.\end{equation}
The mean redshift of the LAB is defined as
\begin{equation}
z = \frac{\lambda_{em} - \lambda_0}{\lambda_0}= 3.0887 \pm 0.0001
\label{eq:zem}
,\end{equation}
where $\lambda_0 = 1215.67$\,\AA\ is the Ly$\alpha$ wavelength at rest (in the vacuum) and
$\lambda_{em}$ is the observed wavelength of the emission corrected to vacuum. The (vacuum)
wavelength $\lambda_{em}= 4970.5 \pm 0.1$\,\AA\ is determined by the first-order moment of the
distribution.
This is a reasonable choice because it takes the skewness in the
shape of the spectrum into account. The redshift of this bright central part of the LAB is therefore
very close to but different from that of the QSO: the difference amounts to $z_p=-0.00154$
or $-462$~km\,s$^{-1}$, about twice the uncertainty on the QSO redshift
(assuming $z_\mathrm{QSO}=3.095\pm 0.003$).

The velocity difference between the PDLA and the LAB is then
\begin{equation*}
\Delta v = v(PDLA)-v(LAB)=-473\pm 24~\mathrm{km\,s}^{-1}\mathrm{,}
\end{equation*}
in perfect agreement with the $-490$~km\,s$^{-1}$ given by \citet{LR99}.

From the redshift, we determine the luminosity distance $d_L$ and the total Ly$\alpha$
luminosity of the blob:
\begin{equation} \label{lum}
L_{Ly\alpha} = 4\pi d_L^2\,F_{Ly\alpha}= (2.18 \pm 0.05)\cdot10^{43} \mathrm{erg\,s}^{-1}
\end{equation}
with $d_L = 26850$\,Mpc. The error given here is a formal error provided by the fit;
a more realistic error would be closer to $\sim 10\%$, taking into account the
uncertainty of the subtracted QSO spectrum.
The FWHM of the average emission spectrum (not corrected
for the instrumental width) is:
\begin{equation} \label{sigma_vel}
FWHM = 9.27 \pm 0.20\,\mathrm{\AA}~~~\mathrm{or}~~~FWHM = 559 \pm 12~\mathrm{km\,s}^{-1}
\end{equation}
if we naively admit that the line width reflects the velocity field alone, without
any transfer effect. This rather modest value already indicates that the gas is not
subject to violent motions, especially as part of the line width may be accounted to
transfer effects, making our FWHM estimate an upper limit rather than a true estimate
of the kinematics.
\citet{LR99} obtain FWHM values varying from $\sim 300$ to $\sim 540$\,km\,s$^{-1}$ (or
even $650$\,km\,s$^{-1}$, but for a line affected by a cosmic ray), which is clearly
smaller. However, their values refer to individual observations of various parts of the
LAB, so they do not include the large scale velocity field (or transfer effects).
They also use a higher spectral resolution, which tends to decrease the measured FWHM,
although this accounts for no more than a difference of  $\sim 25$\,km\,s$^{-1}$ .

\subsubsection{Spatial shape of the central LAB component }
\begin{figure}
 \resizebox{\hsize}{!}{\includegraphics{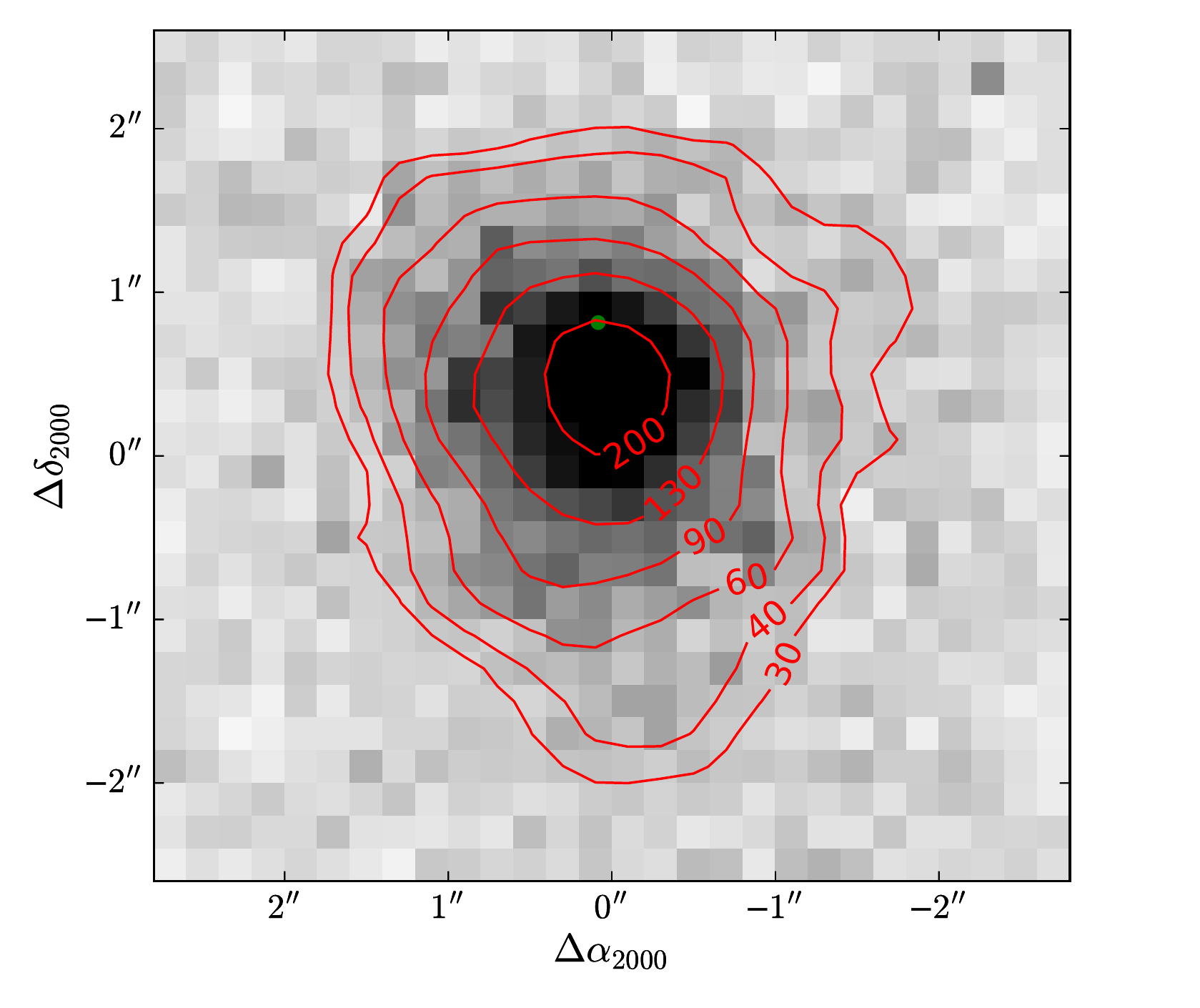}}
 \caption{Image of the LAB obtained by addition of the successive frames from
 $4950$\,\AA\ to $4990$\,\AA. The signal of the QSO has been removed, as
 explained in the text, and the position of the QSO is shown by the green dot.
 The red curves are isophotes spanning SBs from
 $30\cdot10^{-20}$\,erg\,s$^{-1}$\,cm$^{-2}$\,spaxel$^{-1}$ to
 $200\cdot10^{-20}$\,erg\,s$^{-1}$\,cm$^{-2}$\,spaxel$^{-1}$, or
 $7.5\cdot10^{-18}$\,erg\,s$^{-1}$\,cm$^{-2}$\,arcsec$^{-1}$ to
 $5\cdot10^{-17}$\,erg\,s$^{-1}$\,cm$^{-2}$\,arcsec$^{-1}$.
 The outermost isophote corresponds to $S/N\sim 1.8$, while the inner
 corresponds to $S/N\sim 12$ per spaxel.
 There is a slight offset of the QSO position by about $0.4$\arcsec\ with respect to
 the isophotal center of the blob.
 The relative coordinates are defined with respect to the field center.}
 \label{fig:contour}
\end{figure}

Summing the successive layers in the wavelength range of the LAB, we obtain an image of the blob
and can determine its extension and shape. In each spaxel at position $(i,j)$ of the resulting
stacked image we have the sum of all pixel values $I_{ij}(\lambda)$, located at the same position
$(i,j)$, over the interval $D_\lambda$ ranging from $4950$\,\AA\ to $4990$\,\AA\, encompassing
the LAB emission:
\begin{equation}
\bar{I}_{ij} = \sum_{\lambda \in D_\lambda} I_{ij}(\lambda) \Delta\lambda
.\end{equation}
The result is shown in Fig.~\ref{fig:contour}.

The coordinates of the blob center are
\begin{equation}
RA= 21^\mathrm{h}~ 02^\mathrm{m}~ 44.73^\mathrm{s},~~
Dec= -35^\circ~ 53'~ 06.9''~. 
\label{eq:coord_centre_LAB}
\end{equation}
We see that this bright LAB component has a roughly circular shape with an extension of about
$3\arcsec-4\arcsec$.
This corresponds to a diameter of $23-31$\,kpc at the LAB redshift.
\subsubsection{Velocity map of the LAB central part}
To create the velocity map of the blob\footnote{The ``velocity'' word used here is a
proxy for ``line position'' and is strictly justified only when transfer effects can be
neglected.}, we fit a Gaussian profile on each spectrum $\tilde{s}_{ij}$ of
the subcube. From the mean position value of the Gaussian fit, we obtain
the redshift $z$. We wish to obtain the ``peculiar'' redshift $z_p$ at each location of
the blob, however, which has a global redshift $z_{em} = 3.0887$.
\begin{figure*}
\centering
 \includegraphics[width=9cm]{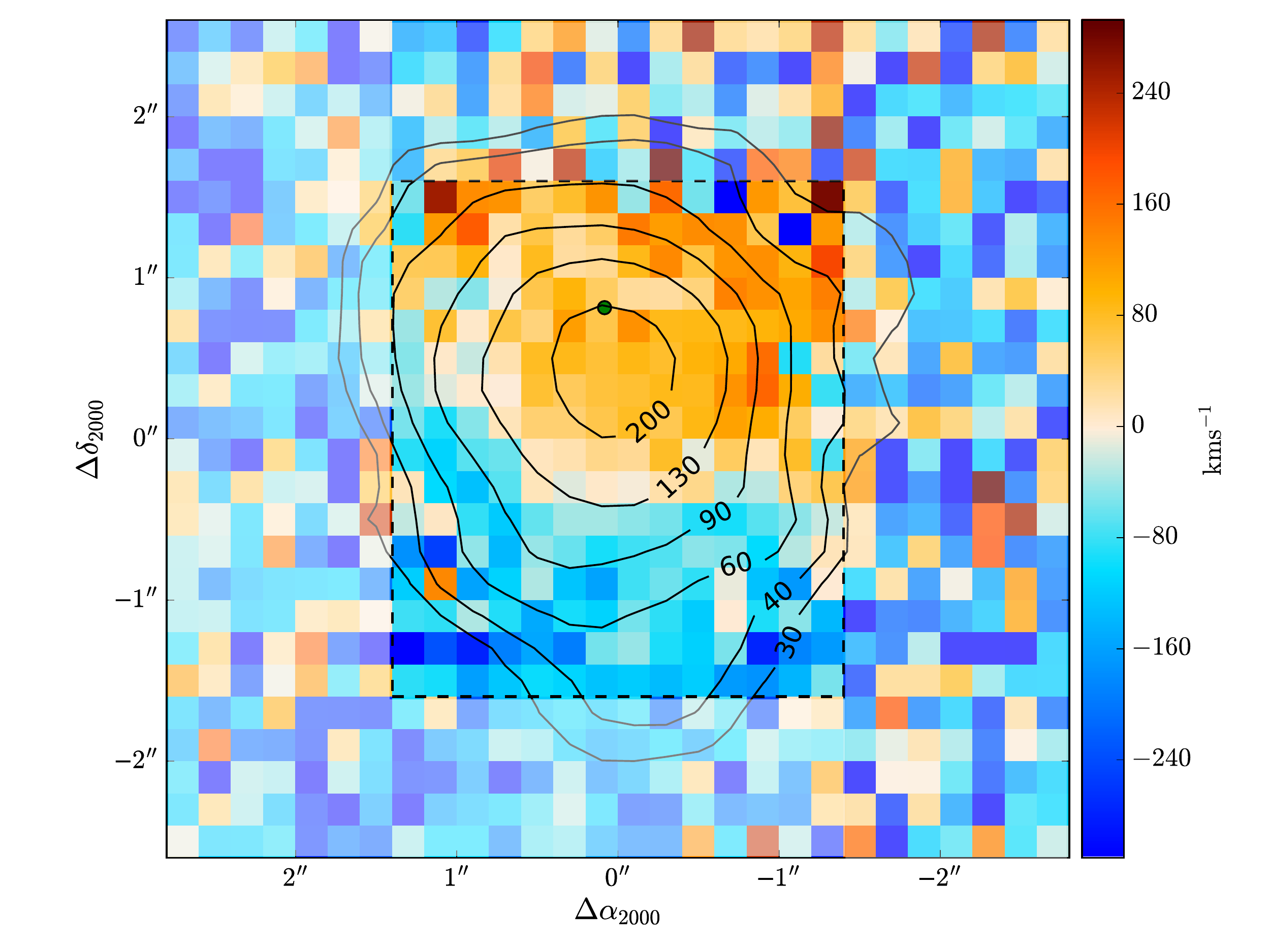}
 \includegraphics[width=9cm]{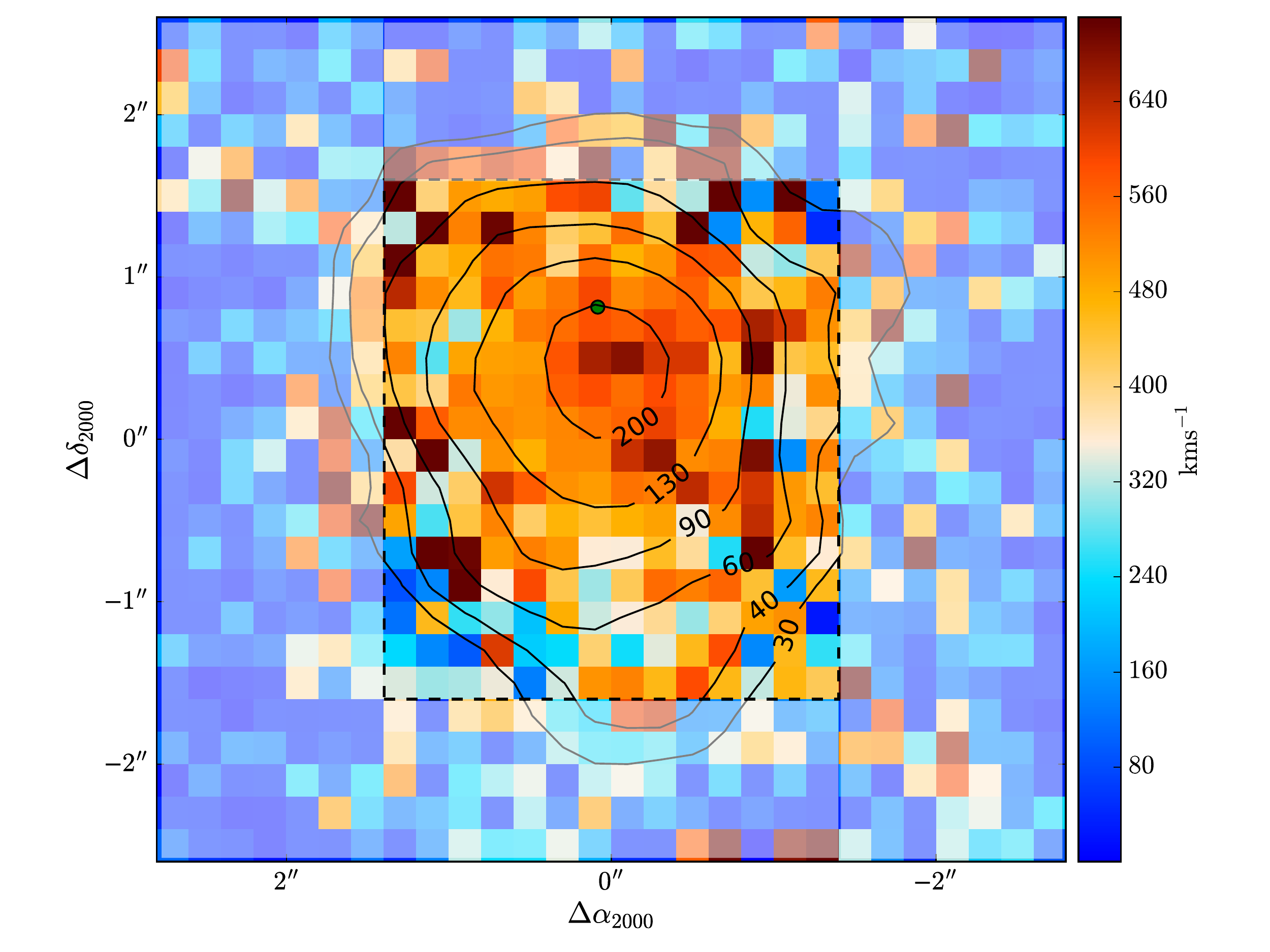}
 \caption{{\sl Left:} Velocity map of the Ly$\alpha$ emission feature. The color code highlights
 the velocity along the line of sight in each spaxel of the image, the zero-point being
 set at $z=3.0887$ or $\lambda_\mathrm{vac}=4970.5$\,\AA. The curves
 are the isophotes labeled in units of $10^{-20}$\,erg\,s$^{-1}$\,cm$^{-2}$\,spaxel$^{-1}$.
 Below $6\cdot10^{-19}$\,erg\,s$^{-1}$\,cm$^{-2}$\,spaxel$^{-1}$ , the S/N is very low and everything
 outside this isophote is strongly affected by background noise.
 The green point near the center shows the position of the QSO.
 The dashed rectangle represents the region of the image used in Fig.~\ref{fig:spectrum_map};
 we have enhanced it by lowering the contrast in the area that surrounds it.
 {\sl Right:} Map of the FWHMs of the Ly$\alpha$ emission feature.
 The FWHM in each spaxel is computed from the Gaussian fit of the emission spectrum.
 For comparison, the average FWHM measured from the total emission spectrum
 is $560$\,km\,s$^{-1}$.}
 \label{fig:vel_map}
\end{figure*}

\begin{figure*}
\centering
 \includegraphics[width=9cm]{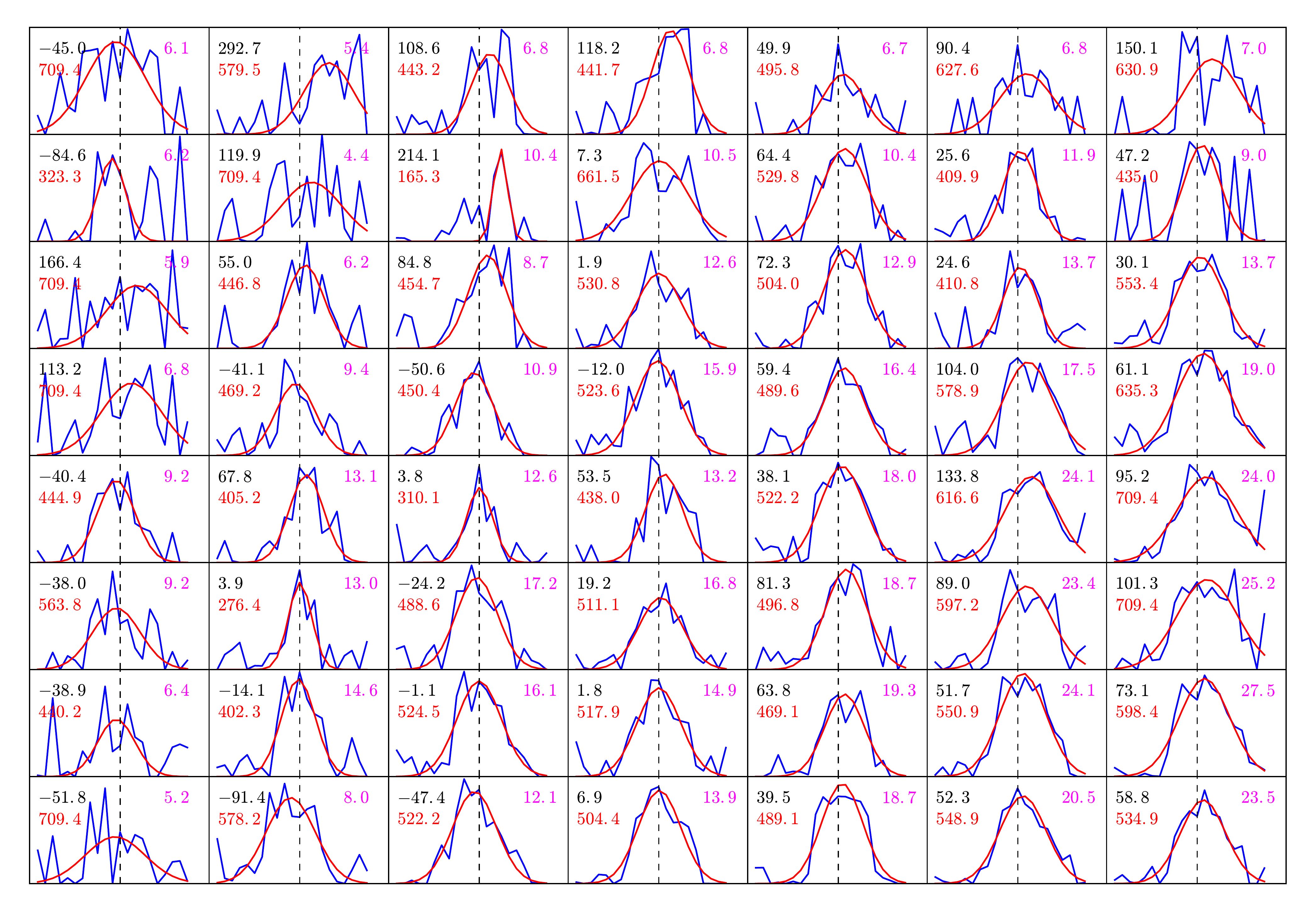}
 \includegraphics[width=9cm]{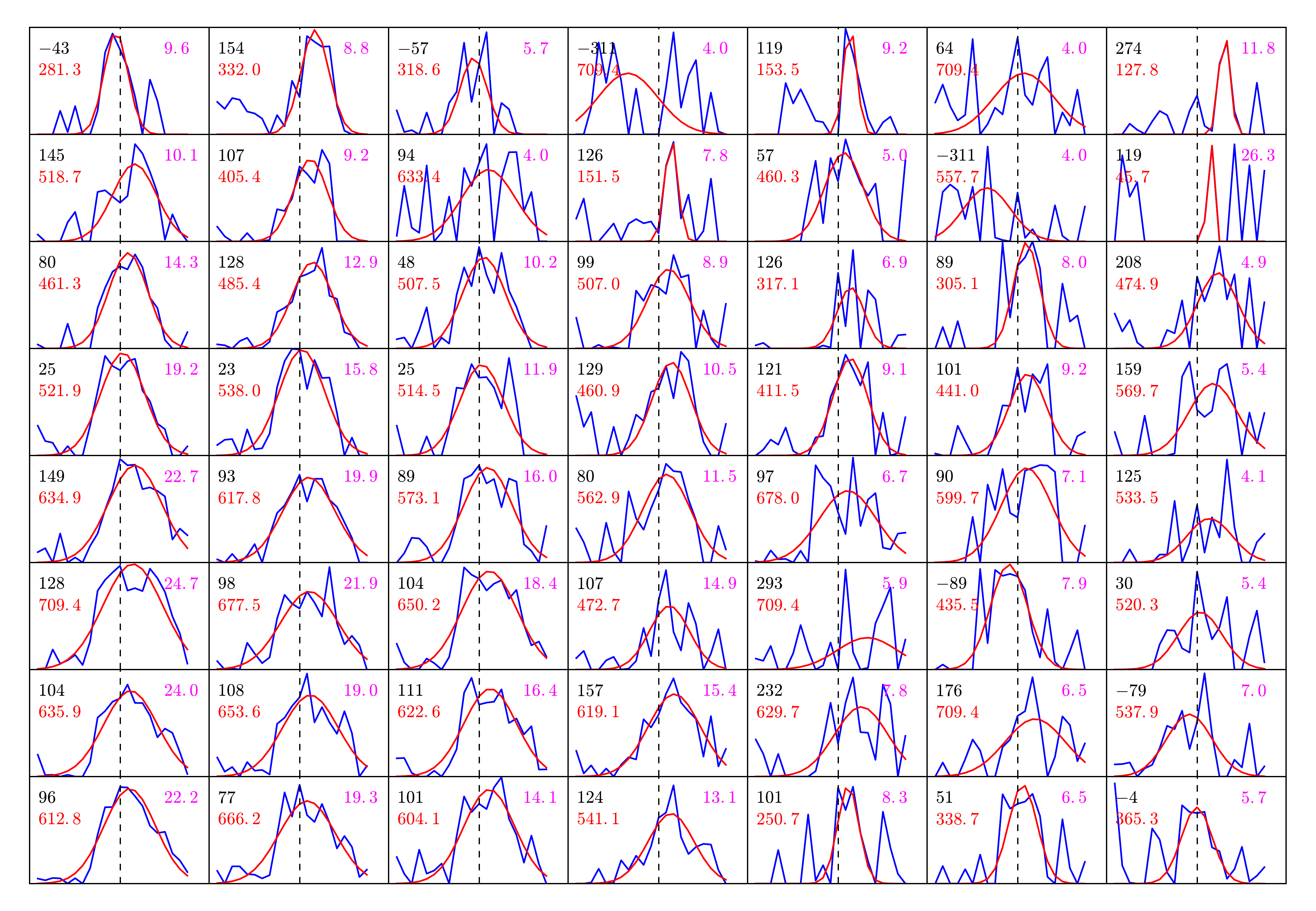}
 \includegraphics[width=9cm]{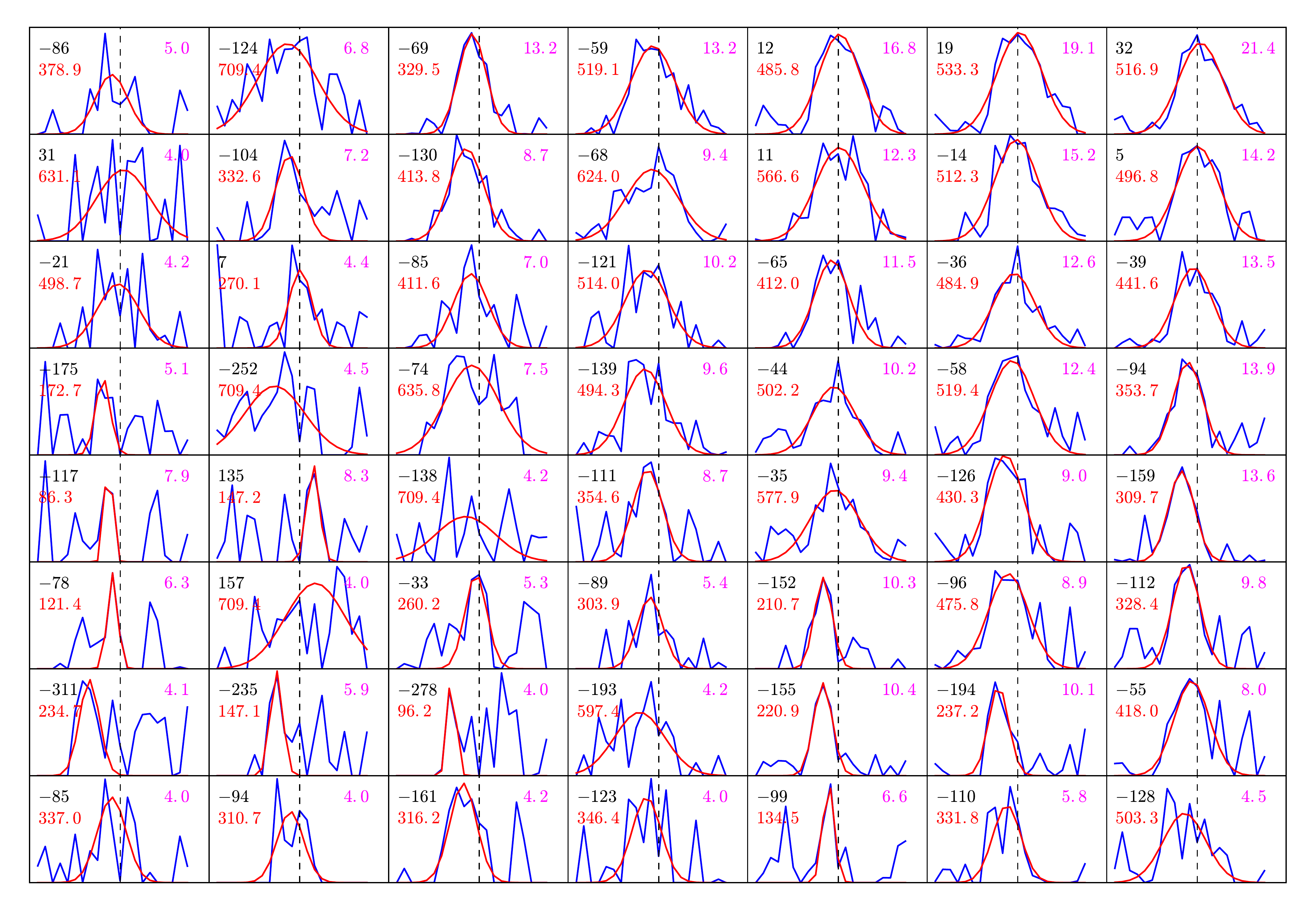}
 \includegraphics[width=9cm]{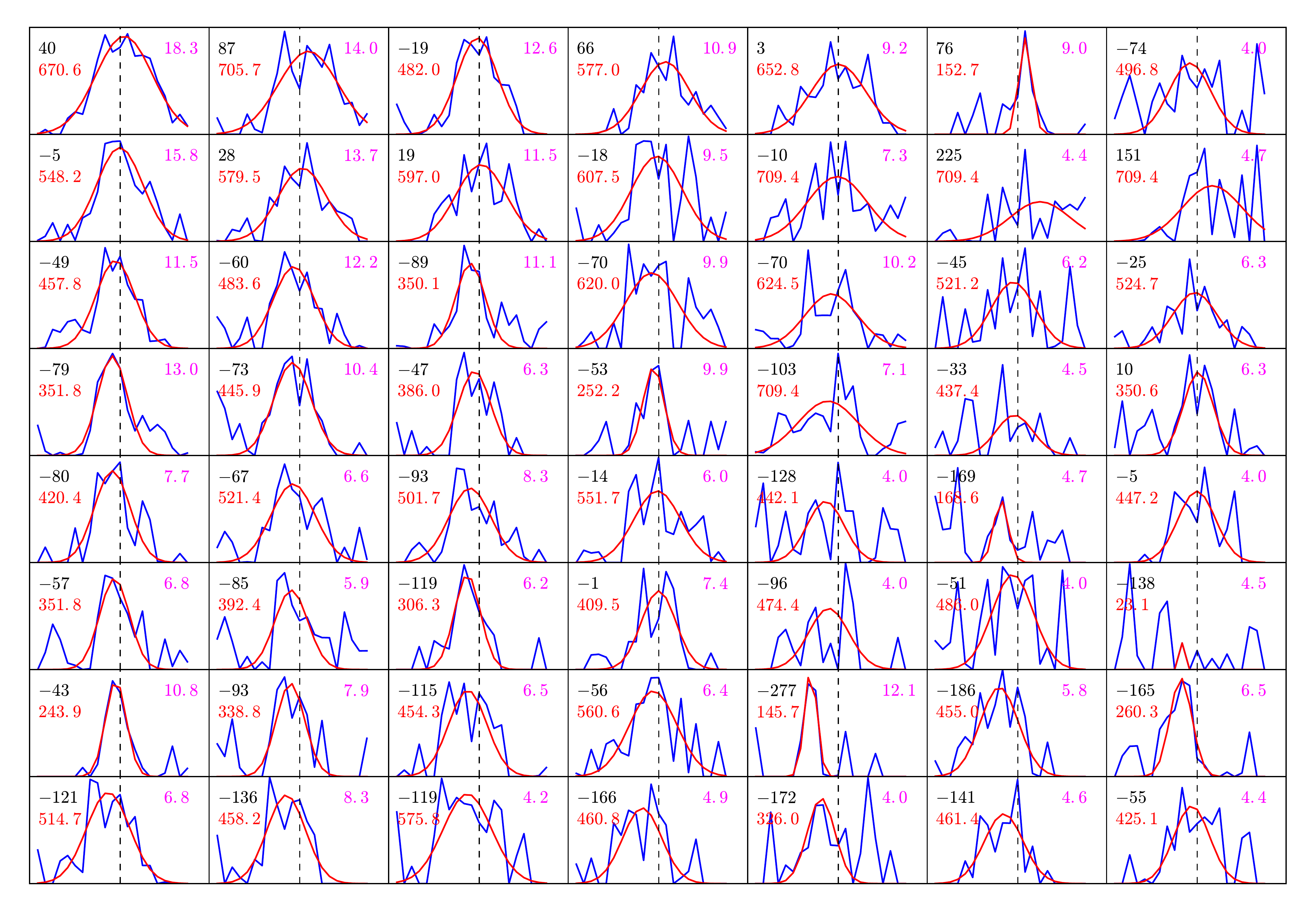}
 \caption{Map of spectra recorded in individual spaxels. The
 total area corresponds to the dashed rectangle in Fig.~\ref{fig:vel_map};
 it covers $14\times16$ spaxels ($2.8$\arcsec$\times 3.2$\arcsec).
 North is up and east to the left.
 The blue lines show the measured spectra, and the red curves
the Gaussian fits.
 The spectral range is $4955$\,\AA$-4980$\,\AA. The black number inside each
 subplot (top left corner) is the peculiar velocity in km\,s$^{-1}$ computed from the
 Gaussian fit; the red number just below is the FWHM
 of the emission line in km\,s$^{-1}$; the purple number (top right) is the amplitude of the
 Gaussian in units of $10^{-20}$\,erg\,s$^{-1}$\,cm$^{-2}$\,\AA$^{-1}$\,spaxel$^{-1}$.
 The dashed vertical line marks the average emission wavelength ($4970.0$\,\AA).
 The quasar lies in the upper left quadrant, in the spaxel [7,5] just next to spaxel [7,4]
 (where [1,1] and [7,8] define the lower left and upper right spaxels).}
 \label{fig:spectrum_map}
\end{figure*}

By applying relations (\ref{eq1}) and (\ref{eq2}) to each spectrum of the subcube, we obtain
the velocity map of the blob, shown in the left panel of Fig.~\ref{fig:vel_map}.
We can clearly distinguish
the presence of two regions, each with a relatively uniform velocity, and separated by
a steep north-south gradient from about
$80$\,km\,s$^{-1}$ at the LAB center, $\sim 0.4$\arcsec\ south of the QSO, to about
$-100$\,km\,s$^{-1}$ $2$\arcsec\ south of the QSO. \citet{LR99}
have shown some evidence for this feature (see their Fig.~4c), but their data
could not provide a complete map such as ours, which also shows an east-west velocity
gradient $\sim 1$\arcsec\ to the east of the LAB center.
We also note that our peculiar velocities overall range from about
$-120$\,km\,s$^{-1}$ to $+80$\,km\,s$^{-1}$. This $\sim200$\,km\,s$^{-1}$
difference is not much larger than our spectral resolution, which is $\sim 163$\,km\,s$^{-1}$,
but it is quite significant. It agrees well with the $\sim 240$\,km\,s$^{-1}$
difference found by \citet[][Table~2 and Fig.~4]{LR99}.

The right panel of Fig.~\ref{fig:vel_map} is a map of the FWHM (obtained through the
Gaussian fit) of the Ly$\alpha$ emission line. When we consider both panels,
a rough overall trend of an increasing FWHM with increasing velocity seems to appear:
the FWHM is around $300$\,km\,s$^{-1}$ $2$\arcsec\ south of the QSO, but closer to
$600$\,km\,s$^{-1}$ at the LAB center. This is also consistent with
\citet[][Fig.~4d]{LR99}.

To verify that the different spectra in each spaxel are well fitted by the Gaussian
functions, we provide Fig.~\ref{fig:spectrum_map}. Each subplot of this figure shows the
total flux recorded in each spaxel. The total area of the figure corresponds
to the dashed rectangle in Fig.~\ref{fig:vel_map}. The Gaussian
fit appears satisfactory in each subplot where the noise is low enough, which occurs
when the fitted amplitude is larger than
$\sim 5\times 10^{-20}$\,erg\,s$^{-1}$\,cm$^{-2}$\,\AA$^{-1}$.

\subsection{Properties of the Lyman alpha blob at larger distance from the QSO and other objects}\label{large}
\begin{figure}
 \resizebox{\hsize}{!}{\includegraphics{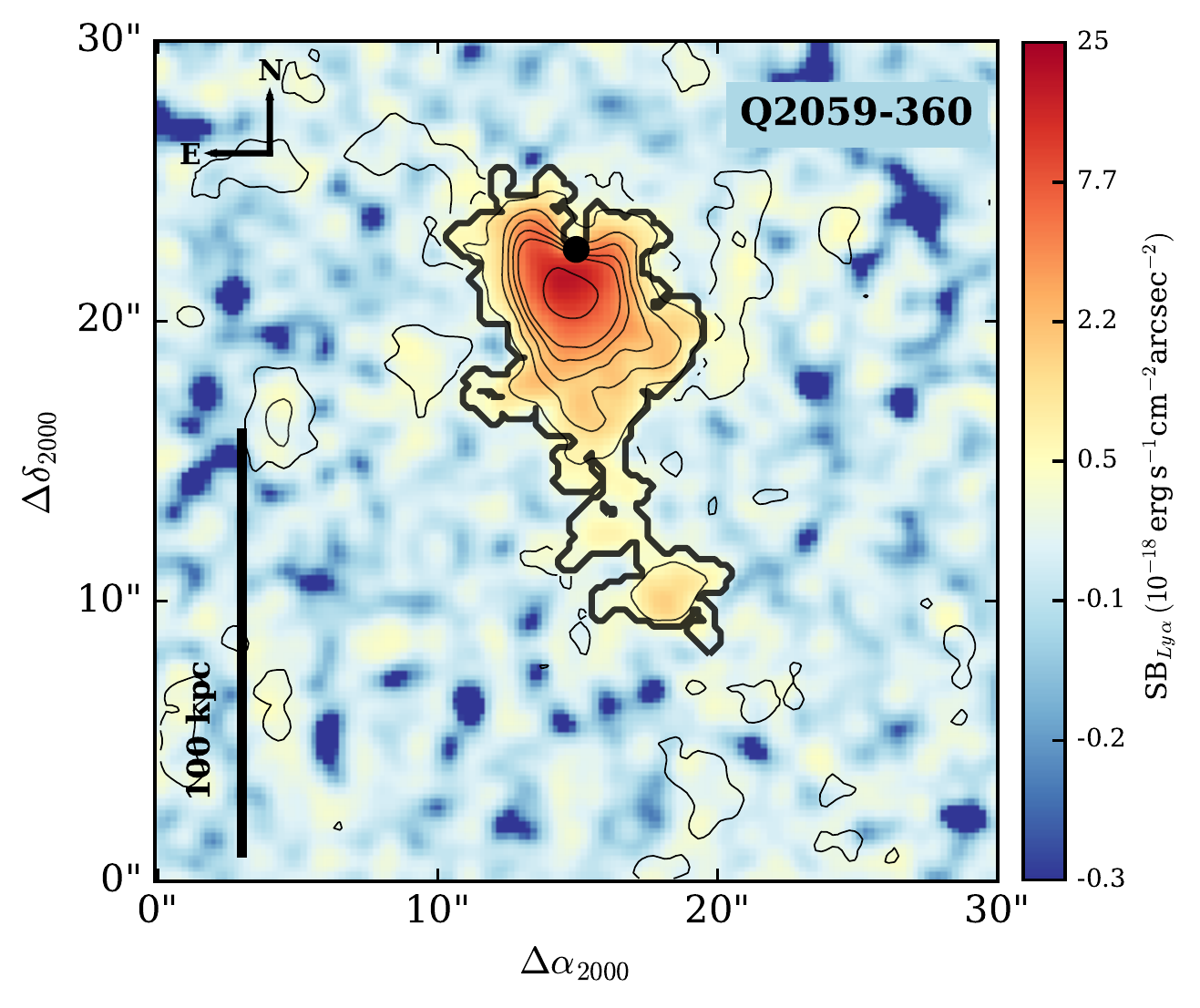}}
 \caption{Optimally extracted map of the LAB, obtained using the CubEx software
 on the PSF and continuum subtracted MUSE datacube, and by collapsing along the wavelength
 axis the datacube voxels inside the 3D segmentation map. The spatial projection of the
 3D mask is indicated by the thick contour that corresponds to an SB of about
 $1\times 10^{-18}$\,erg\,s$^{-1}$\,cm$^{-2}$\,arcsec$^{-2}$ and an S/N of $2.3$.
 The thin contours indicate the propagated S/N in the images, the outermost
 one corresponding to S/N$=2$ and the other ones to
 S/Ns increasing by steps of $6$ units (see B16 for a
 detailed explanation). The QSO position is shown by the black dot.
 Note the long tail extending southward of the QSO; this was unknown to date.}
 \label{fig:map_LAB}
\end{figure}

Using the CubExtractor software, we were able to map the whole LAB, especially at large distance
from the QSO, as seen in Fig.~\ref{fig:map_LAB}. The LAB extends over $100$\,pkpc at least,
like the 17 other LABs observed around radio-quiet QSOs by B16. The large
extension is mainly due to a filament protruding southward and the Lya flux measured
within the 3D mask (thick contour) is $1.67\times 10^{-16}$\,erg\,s$^{-1}$\,cm$^{-2}$.
This is lower than what we found for the bright part of the LAB, probably because of the poor
subtraction of the QSO light at the QSO position, which artificially reduces the
LAB intensity there.
The filament is relatively thin but appears resolved, being wider than $1$\arcsec\ or $8$\,pkpc.

The morphology of this LAB is quite similar to that of several LABs discovered by
\citet{MYH11} through the Subaru Ly$\alpha$ blob survey in the SSA22 area at the same
redshift. In particular, SSA22-Sb6-LAB4 has almost exactly the same shape and extent,
with a thin filament protruding to the southwest. The only difference is the absence
of any obvious QSO.

\begin{figure*}
 \includegraphics[width=9.4cm]{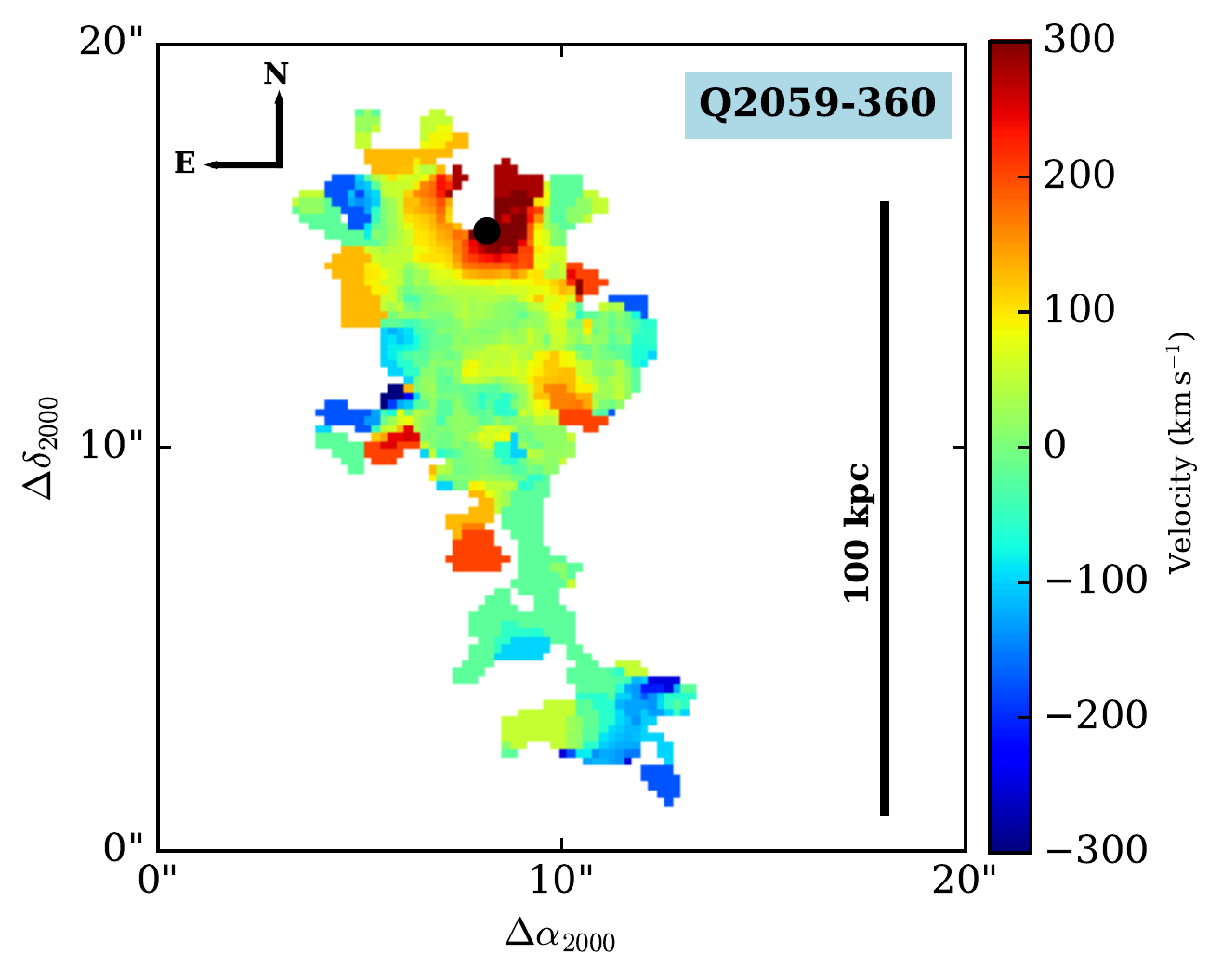}
 \includegraphics[width=9.0cm]{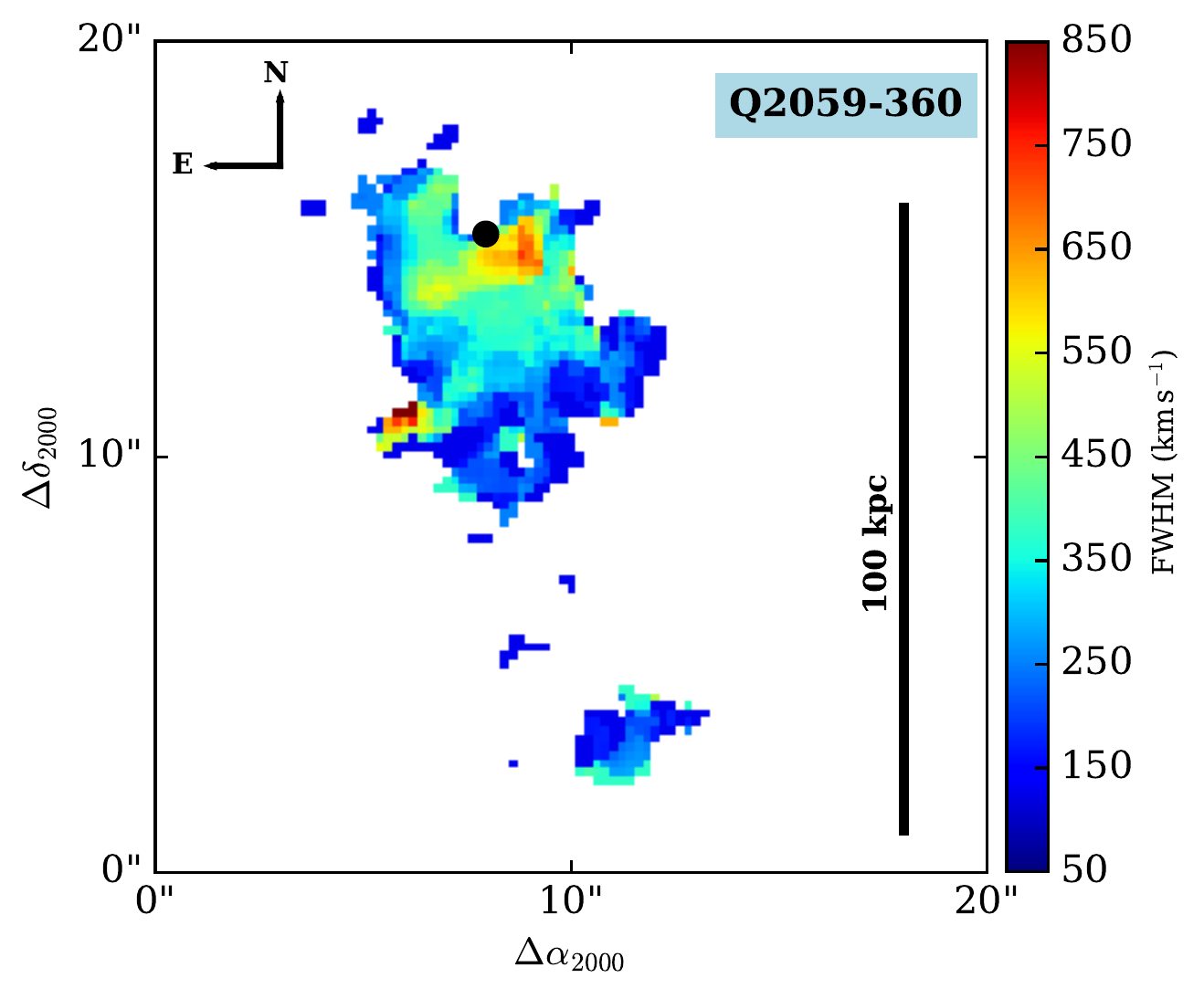}
 \caption{{\sl Left:} Large-scale velocity map of the LAB given by the CubEx software on
 the basis of the first moment of the Ly$\alpha$ emission line. The QSO position is
 marked by the black dot. The zero velocity corresponds to
 $\lambda_\mathrm{vac}=4967.26$\,\AA, or to $-196$\,km\,s$^{-1}$ in Fig.~\ref{fig:vel_map}
 (left panel). {\sl Right:} Large-scale map of the velocity dispersion in the LAB
 given by the CubEx
 software on the basis of the second moment of the Ly$\alpha$ emission line.}
 \label{fig:vel_map_wide}
\end{figure*}

The corresponding velocity map is shown in the left panel of Fig.~\ref{fig:vel_map_wide},
the zero velocity corresponding to $\lambda_\mathrm{vac}=4967.26$\,\AA\ (maximum of the
emission intensity), that is, to $-196$\,km\,s$^{-1}$ in Fig.~\ref{fig:vel_map}
(left panel). It is based on
the first moment of the emission line, while the velocity map of the immediate vicinity
of the QSO (Fig.~\ref{fig:vel_map}) is obtained through Gaussian fits. This might partly explain the larger velocity contrast ($\sim 300$\,km\,s$^{-1}$ instead of $\sim 200$\,km\,s$^{-1}$)
near the QSO. Another cause for the discrepancy may be the approximate subtraction
of the QSO PSF. Nevertheless, the two velocity maps agree regarding the north--south
velocity gradient, which appears very steep about $1$\arcsec\ south of the QSO.

At larger scale, the southern component
of the LAB overall has a velocity close to $0$\,km\,s$^{-1}$, while in the close
QSO neighborhood, the velocity is higher than $\sim 100$\,km\,s$^{-1}$.

The map of velocity dispersion (as a proxy for the line width) is shown in the right panel
of Fig.~\ref{fig:vel_map_wide}. We note the very small dispersion in the southern component
of the LAB, which is essentially equal to the MUSE resolution limit. The same is true of the 
southern part of the main LAB component. Near the QSO, the Ly$\alpha$ line is broader on an oblique
band stretching from the SW of the QSO to the SE of it. Our previous map (Fig.~\ref{fig:vel_map},
right panel)
does not show this feature so clearly, but does display widths between $\sim 240$ and
$\sim 600$\,km\,s$^{-1}$, which are similar.
There is rough agreement between the two maps in the sense that the Ly$\alpha$ line
tends to be broader near the QSO than far away from it.

\begin{figure}
 \resizebox{\hsize}{!}{\includegraphics{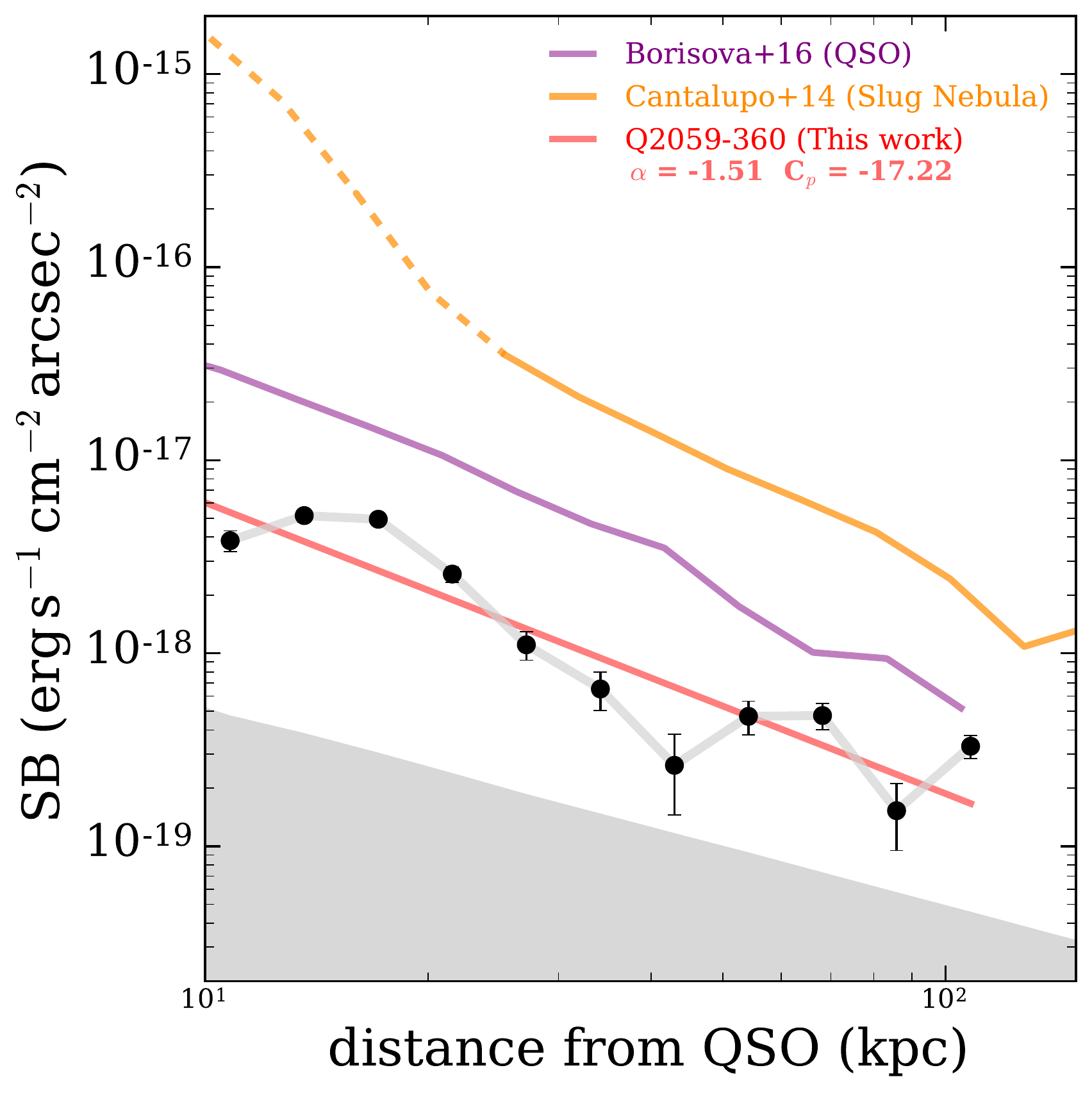}}
 \caption{Radial variation of the SB of the LAB (black dots; distance
 in physical kpc). The red line
 is a least-squares fit of a power law to the data. The average profile of the
 $17$ nebulae associated with radio-quiet QSOs (B16) is shown as a purple line.
 The yellow line represents the Slug Nebula \citep{CAP14}, the dashed part of it is
 polluted by the QSO light.
 Correcting the observed SB for redshift dimming would bring the yellow curve very
 close to the purple curve, but would slightly increase the distance between the red and
 purple curves.
 The upper limit of the gray area corresponds to the $2\sigma$
 noise on the profile.}
 \label{fig:SB_radial}
\end{figure}

\subsubsection{Surface brightness profile and size--luminosity relation}
We show the mean radial variation of the azimuthally averaged surface brightness (SB) in 
Fig.~\ref{fig:SB_radial},
as measured on the SB map of Fig.~\ref{fig:map_LAB}. The average is taken over circular
annuli.
Although the LAB is strongly asymmetric, it is worth comparing this radial variation
with that of other blobs. As mentioned above, the innermost part (radius $\lesssim 1$\arcsec)
is not reliable, but the rest is, and the SB curve shown here was obtained in exactly
the same way as the $17$ others published by B16. It is essentially free of
any pollution from the QSO light because it shows only the result beyond
$10$\,kpc of it, that is, beyond $1.3$\arcsec. The red line shows a power-law fit
(see B16, Appendix B) with an index $\alpha=-1.51$, slightly lower than the average
shown by the purple line and corresponding to $\alpha=-1.8$. Only four of the 19 objects
(including two nebulae associated with radio loud QSOs) of B16 have a shallower
profile. The exceptionally
elongated and filamentary shape of our LAB might be responsible for this unusual profile.

Figure~\ref{fig:SB_radial} also shows the SB profile of
the Slug Nebula
\citep{CAP14}. We see that in the range of radii considered,
the profile of our LAB lies almost
an order of magnitude below the average of the B16 sample. Overall, our LAB compares well with
the sample of B16, even though it lies on the faint side.

This is also confirmed by the relation between luminosity and size shown in
Fig.~\ref{fig:lum_size}. The size is defined here as the major axis of the nebula.
The objects observed by B16 are represented as blue stars, while the full magenta
dots are the LABs discovered by \citet{MYH11}. Interestingly, the latter largely
overlap the former, but extend to lower luminosities and smaller sizes. This is
probably an observational bias: B16 have selected the brightest QSOs, so one may
expect that the luminosity of the associated LABs is enhanced by fluorescence due
to the QSO UV radiation. On the other hand, Matsuda's LABs do not harbor any obvious
QSO. Thus, our object indeed lies on the faint end of the LABs surrounding
bright QSOs, but appears to be a rather average one when compared to the general LAB
population, as far as the sample of \citet{MYH11} is representative.

\subsection{Upper limits to high-ionization lines}
\begin{figure}
 \resizebox{\hsize}{!}{\includegraphics{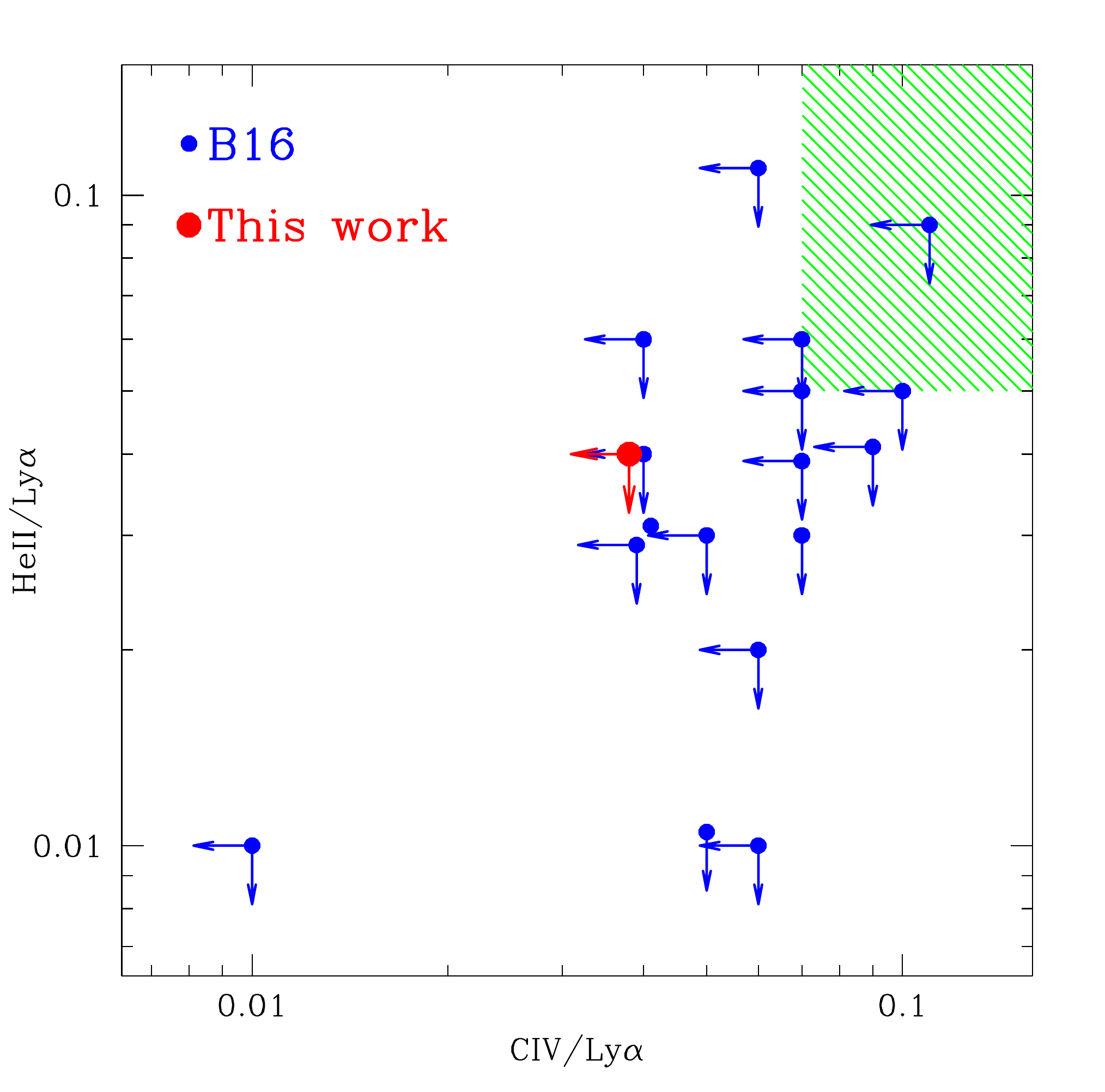}}
 \caption{Ratios of high-ionization line fluxes for our object (red) and for those
 of B16 (Table~2) surrounding radio-quiet QSOs (blue). The \ion{He}{ii} line is
 detected in only one object of the latter sample, while the \ion{C}{iv} line is
 detected in three$\text{}$ objects. The green hatched area represents the range of line
 ratios compatible with photoionization models \citep{ABY15}.}
 \label{fig:HeII}
\end{figure}

\begin{figure}
 \resizebox{\hsize}{!}{\includegraphics{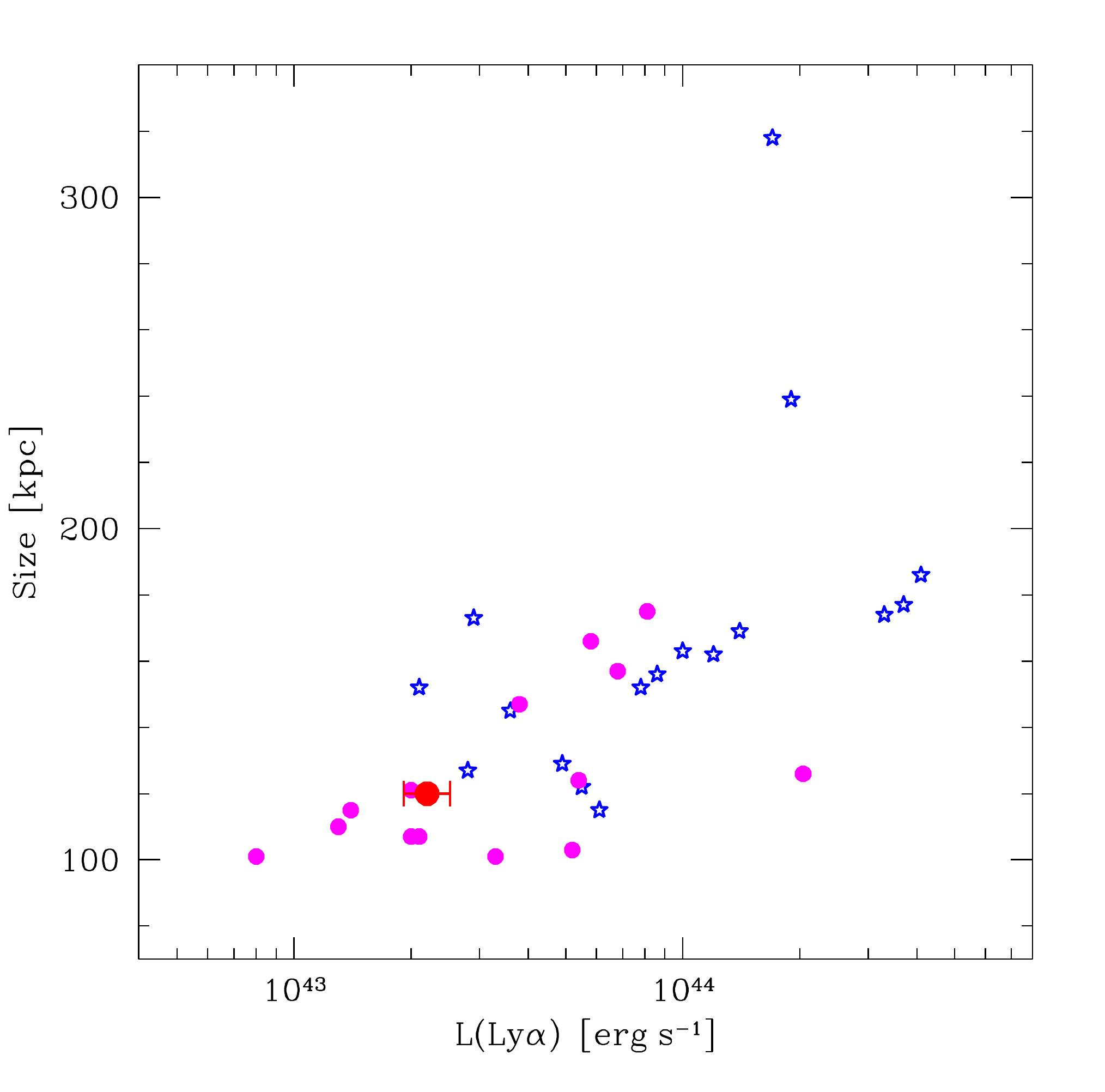}}
 \caption{Relation size--luminosity for our object (red) and for those
 of B16 that are radio quiet (blue). The magenta full dots represent the
 14 giant LABs discovered by \citet{MYH11}.}
 \label{fig:lum_size}
\end{figure}

We examined the QSO-subtracted datacube for any presence of the
\ion{N}{V}$\lambda 1238.8-1242.8$, \ion{C}{IV}$\lambda 1548.2-1550.8$, \ion{He}{II}$\lambda 1640.4$,
and \ion{C}{III]}$\lambda 1908.7$. We defined $2\sigma$ upper limits to the flux
of these lines, using the same method as in B16, and compared them with the flux
of the Ly$\alpha$ line defined in the same datacube. The results are given in
Table~\ref{tab:high_ion}, and Fig.~\ref{fig:HeII} shows the ratio of the \ion{He}{II}
and Ly$\alpha$ fluxes versus the
ratio of the \ion{C}{IV} and Ly$\alpha$ fluxes for our object and for the B16 sample.
Here again, our object appears quite similar to the other nebulae found around radio quiet
QSOs.
\begin{table}
\caption{Upper limits to the SB of lines due to highly ionized species.} 
\label{tab:high_ion}
\centering
\begin{tabular}{lcc}
Line, $\lambda$ (vac.)& Flux ($2\sigma$ limit)  & $\frac{Flux(line)}{F(Ly\alpha)}$ \\ 
\multicolumn{1}{c}{[\AA]} &[$10^{-18}$\,erg/s/cm$^{2}$/arcsec$^{2}$]&          \\ \hline\\
\ion{N}{V}\,$1238.8-1242.8$ &$<10.3$&$0.058$\\
\ion{C}{IV}\,$1548.2-1550.8$&$< 6.7$&$0.038$\\
\ion{He}{II}\,$1640.4$      &$< 7.1$&$0.040$\\
\ion{C}{III]}\,$1908.7$     &$<22.1$&$0.125$\\ \hline
\end{tabular}
\end{table}
\citet{ABY15} have shown that in order to rule out the photoionization scenario, one
would need to reach upper limits as low as $0.05$ and $0.07$ on the \ion{He}{II}/Ly$\alpha$
and \ion{C}{IV}/Ly$\alpha$ ratios, respectively. This is indicated by the green area in
Fig.~\ref{fig:HeII}: photoionization as modeled by \citet{ABY15} is possible within that area,
but would imply unrealistic hydrogen column densities ($>10^{22}$~cm$^{-2}$) beyond it
(the authors assume spherical clouds of cool gas).
None of the high-ionization lines
available in the MUSE spectral range can be seen, and our upper limit on the
\ion{C}{iv}/Ly$\alpha$
flux ratio seems compelling enough to exclude photoionization by the QSO as the main driver
of the LAB Ly$\alpha$ luminosity. The upper limit on the \ion{He}{II}/Ly$\alpha$ ratio is just
below the border mentioned, so that based on this criterion alone, our object would
be on the verge of being compatible with
photoionization by the central source. Interestingly, only one among the B16
radio quiet targets lies in the photoionization region in Fig.~\ref{fig:HeII}, but since its
position corresponds to upper limits only, it remains impossible to tell how far its luminosity
can really be explained by photoionization. All other objects have a too low line ratio
along either one or both axes.

\subsubsection{Two other probable Lyman alpha emitters at the
same redshift}\label{other}
We have detected two other possible Ly$\alpha$ emission features in the neighborhood of the
QSO; their coordinates are given in Table~\ref{table:LAE_prop}.

\begin{figure*}
 \includegraphics[width=9cm]{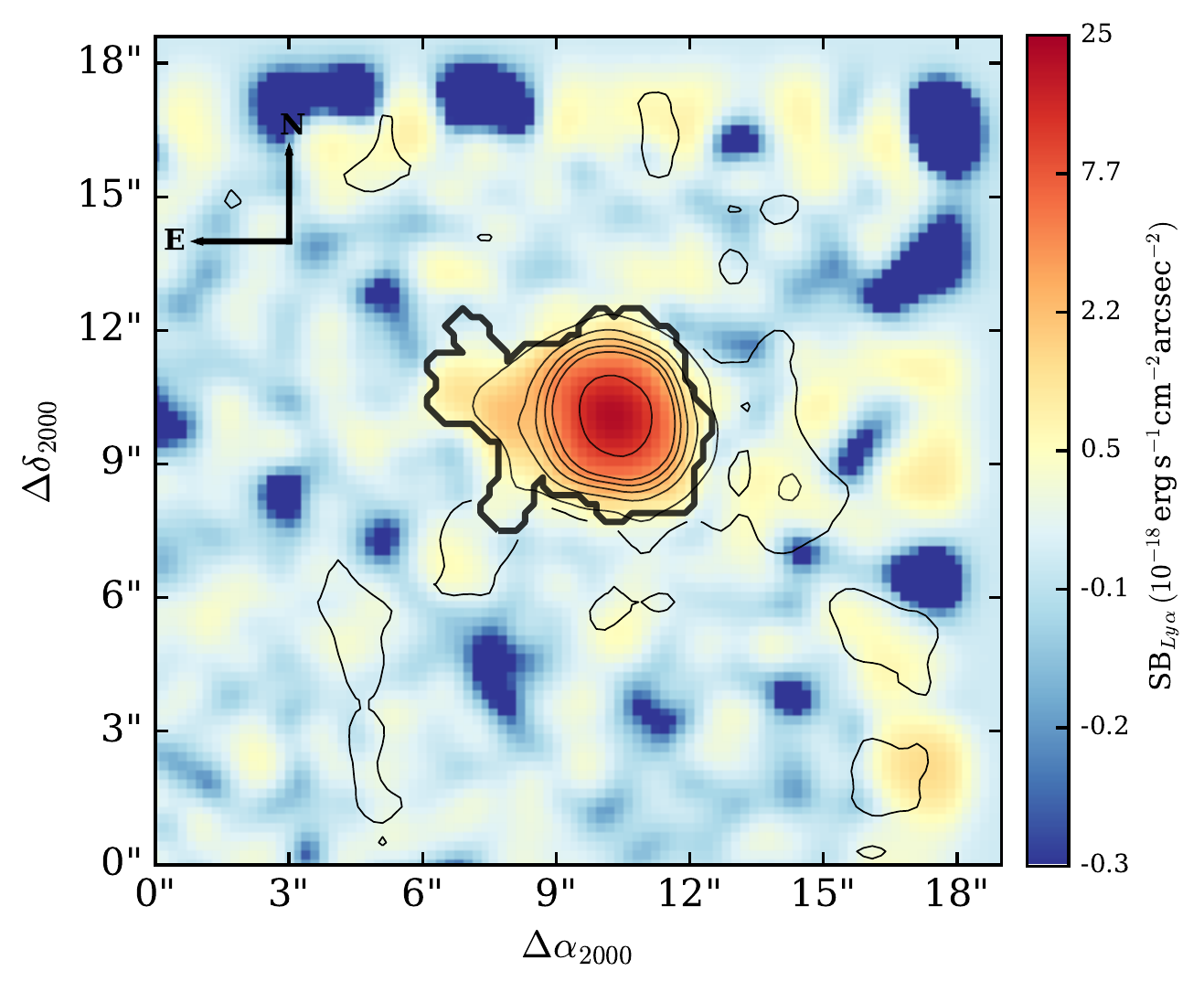}
 \includegraphics[width=9.3cm]{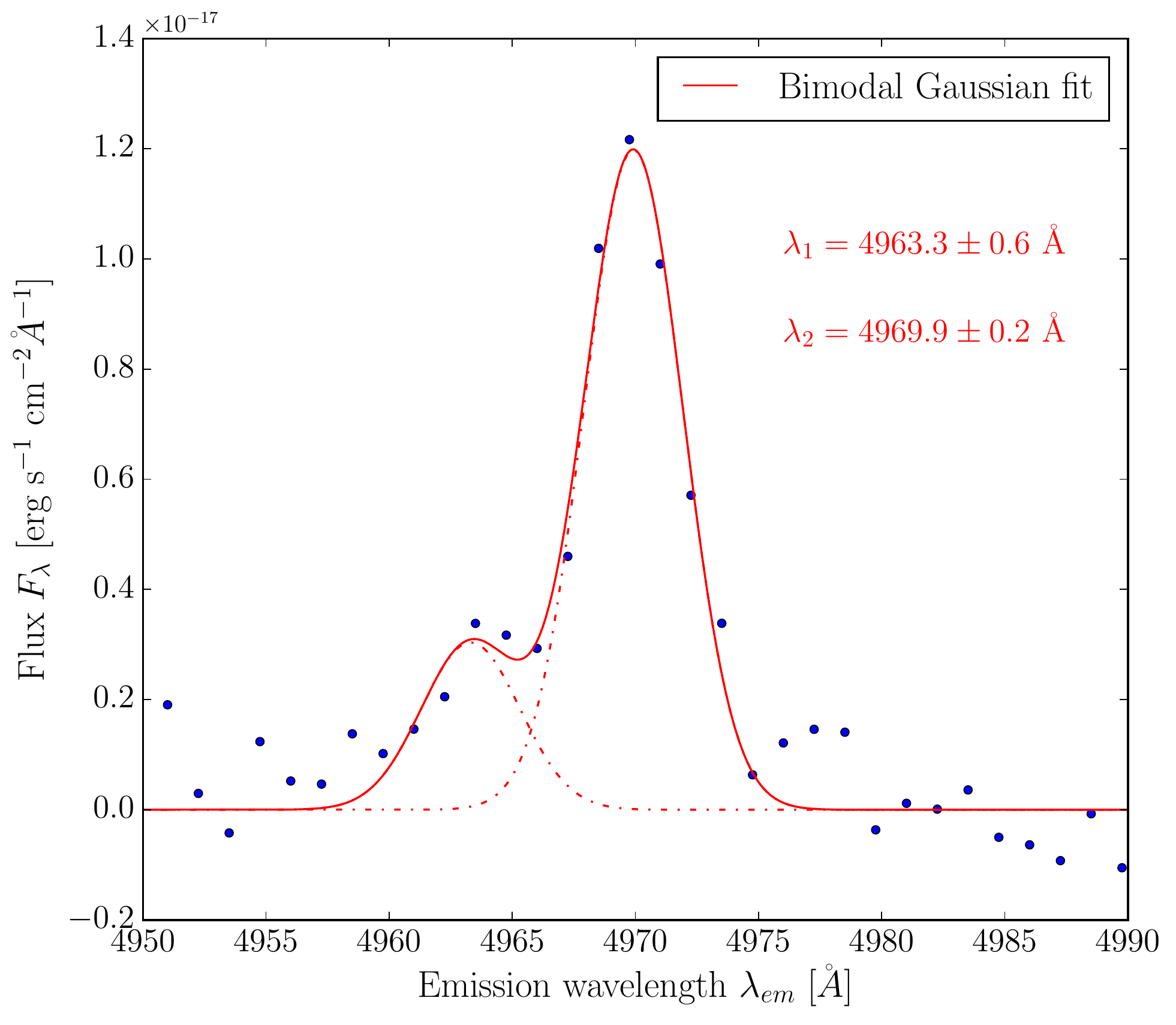}
 \caption{{\sl Left:} CubEx optimally extracted map of the probable Ly$\alpha$ emitter
 LAE1 located $34$\arcsec\ away from
 the QSO. The key to the curves is the same as for Fig.~\ref{fig:map_LAB}.
 {\sl Right:} Spectrum of the LAE1 object, summed on a disk with $2.0$\arcsec\ radius.
 The broad continuous curve shows a fit performed with a
 bimodal Gaussian functions. The continuum has been subtracted and the wavelengths
 are corrected to vacuum.}
 \label{fig:LAE1}
\end{figure*}

The brighter feature (LAE1) lies $34$\arcsec\ away from the QSO, which corresponds
to a projected distance of about $265$~pkpc, and it emits within the range $4960$\,\AA\ to $4975$\,\AA.
The vicinity to the QSO and the emission at wavelengths identical to the central LAB
suggest that this object is emitting in the Ly$\alpha$ line.
Figure~\ref{fig:LAE1} (left panel) shows the extension of this emitter as found with CubEx,
which is about $4.3$\arcsec\ or $34$~pkpc in diameter; if the eastern extension is real
and physically belongs to the LAE, then the major axis of the LAE would reach
$6.4$\arcsec\ or $50$~pkpc. This appears very large, but is consistent with the
finding of \citet{WBB16}, who found many examples of LAEs with Ly$\alpha$ emission
extending to $2$\arcsec\ radius at similar redshift.

The emission feature of this nebula consists of two peaks, as shown in the right
panel of Fig.~\ref{fig:LAE1}.
This is very typical of some Ly$\alpha$ emitters (LAEs). LAE 221745.3+002006, for instance,
shows a very similar profile \citep[Fig.~2]{YMK12}. The latter authors concluded
from their observation of $91$ LAEs that more than $40$\% of them show a double peak,
and that in most cases the red component is stronger than the blue one, which is naturally
explained by transfer effects in expanding media, even inhomogeneous ones
\citep{VSM06,VSA08,VDB12,GD16}.

Its total Ly$\alpha$ flux is
\begin{equation*}
F_{Ly\alpha} = (7.6 \pm 0.4)\cdot10^{-17} \textnormal{ erg s$^{-1}$ cm$^{-2}$}
\end{equation*}
and the positions and widths of the two peaks are summarized in Table~\ref{table:LAE_prop}.
\begin{table*}
\caption{Equatorial coordinates of the two LAEs detected at same redshift as the LAB.
The central positions $\lambda$ (corrected to vacuum) and FWHM
of the two Gaussian fits to each LAE Ly$\alpha$ emission line are given as well.
The $\Delta v$ quantity is the velocity difference
between the peak position and the first moment of the average Ly$\alpha$ profile of the
LAB surrounding the QSO.} 
\label{table:LAE_prop}
\centering
\begin{tabular}{ccclccccc}
Object&\multicolumn{2}{c}{$\alpha$~~~~~~(J2000)~~~~~~$\delta$}&peak&$\lambda$&z&\multicolumn{2}{c}{\small{FWHM}}&$\Delta v$ \\ 
      &                          [h:m:s]    & [\degr:':"] &     & [\AA]   & &   [km\,s$^{-1}$]&[$\AA$]&[km\,s$^{-1}$]\\
      \hline\\
LAE1  &$21:02:42.711$&$-35:52:42.7$&blue  &$4963.3\pm 0.6$&$3.0828\pm 0.0005$&$284 \pm 90$ & $4.7 \pm 1.5$ &$-434\pm 36$\\
      &              &             &red   &$4969.9\pm 0.2$&$3.0882\pm 0.0001$&$287 \pm 23$ & $4.8 \pm 0.4$ &$ -36\pm 12$\\
      \hline\\
LAE2  &$21:02:43.896$&$-35:52:42.6$&blue  &$4963.4\pm 0.8$&$3.0828\pm 0.0006$    &$284 \pm 112$& $4.7\pm 1.9$ &$-429\pm 48$\\
      &              &             &red   &$4972.9\pm 0.3$&$3.0906\pm 0.0002$    &$348 \pm 44 $& $5.8\pm 0.7$ &$ 145\pm 18$\\ \hline
\end{tabular}
\end{table*}
The emission wavelength of the red peak coincides closely (i.e., within $0.6$~\AA\ or
$36$~km\,s$^{-1}$) with the first moment of the emission
line of the LAB surrounding the QSO. Thus, assuming that the double-peak appearance is due
to transfer in an expanding medium, the unabsorbed line would be centered at about $-230$\,km\,s$^{-1}$
with respect to the LAB emission. Such a small velocity difference suggests that this LAE
is indeed close to the QSO+LAB system.

At this point, the question arises whether LAE1 might owe at least part of its luminosity to
photoionization by the QSO (fluorescence scenario). In this case, even in the most extreme
case of a purely external photoionizing source, a double peak similar to the observed peak
would be produced \citep{CPL05,VOS15}, so the answer is positive.

\citet{YMK12} observed $91$ LAEs with a spectral
resolution $R\simeq 1700$, very similar to that of MUSE. They show in their Fig.~5 the
empirical relation
between the velocity separation between the red and blue peaks and the FWHM of the stronger
peak. Our LAE1 object has a rather small velocity separation ($392\pm 36$\,km\,s$^{-1}$) for the
width of the stronger (red) peak, sending its representative point onto the lower envelope
defined by the objects of \citet{YMK12}. It is also reminiscent of the Lyman continuum
emitters (LCEs) examined by \citet{VOS17}.

Using Eq.~\ref{lum} and the luminosity distance corresponding to the average redshift $z=3.0855$,
we obtain the luminosity
\begin{equation*}
L_{Ly\alpha} = (6.5 \pm 0.3)\cdot10^{42} \textnormal{ erg\,s$^{-1}$}
\end{equation*}
We do not show any velocity or FWHM map for this emitter because of its small spatial extension.

In a white-light image (obtained by integrating the datacube over all wavelengths), we see
a very faint object at the position of the Ly$\alpha$ emission, which confirms that we
are not in the extreme case of the ``dark galaxies'' seen by \citet{CLH12}. The rest-frame
equivalent width ($EW_0$) obtained using the same $\text{five}$-pixels radius aperture for both
the Ly$\alpha$ line and the continuum (broad band --- $2000$\,\AA\ --- image reconstructed
from the MUSE datacube and assuming a $\beta$ slope of $-2$) is $EW_0 < 130$\,\AA,
while a dark galaxy is defined by $EW_0 > 240$\,\AA. Therefore, LAE1 must be powered
from inside, for instance, by some stellar formation, although we cannot exclude some contribution
of fluorescence due to the QSO, as mentioned above.

\begin{figure*}
 \includegraphics[width=9.5cm]{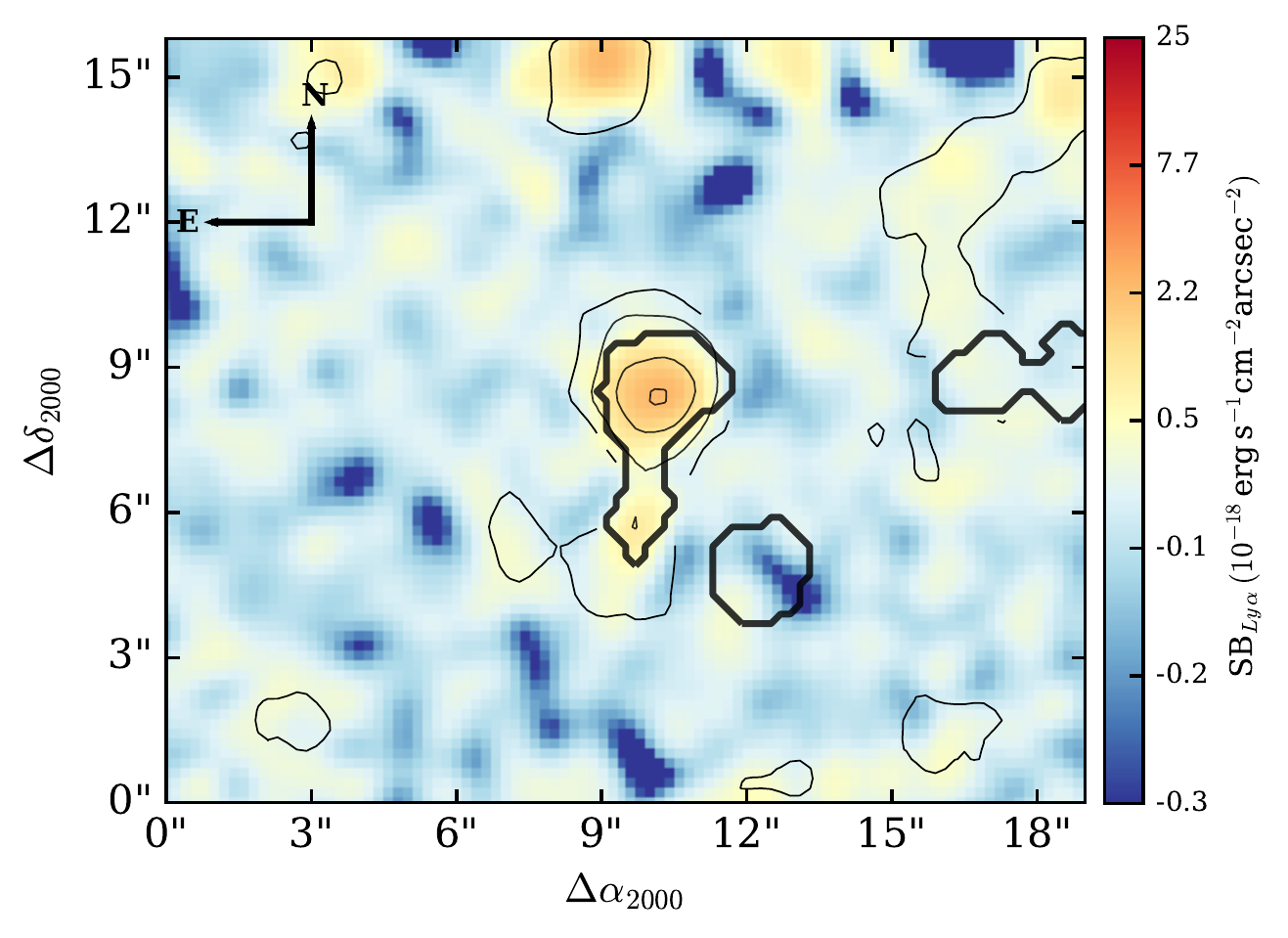}
 \includegraphics[width=8.5cm]{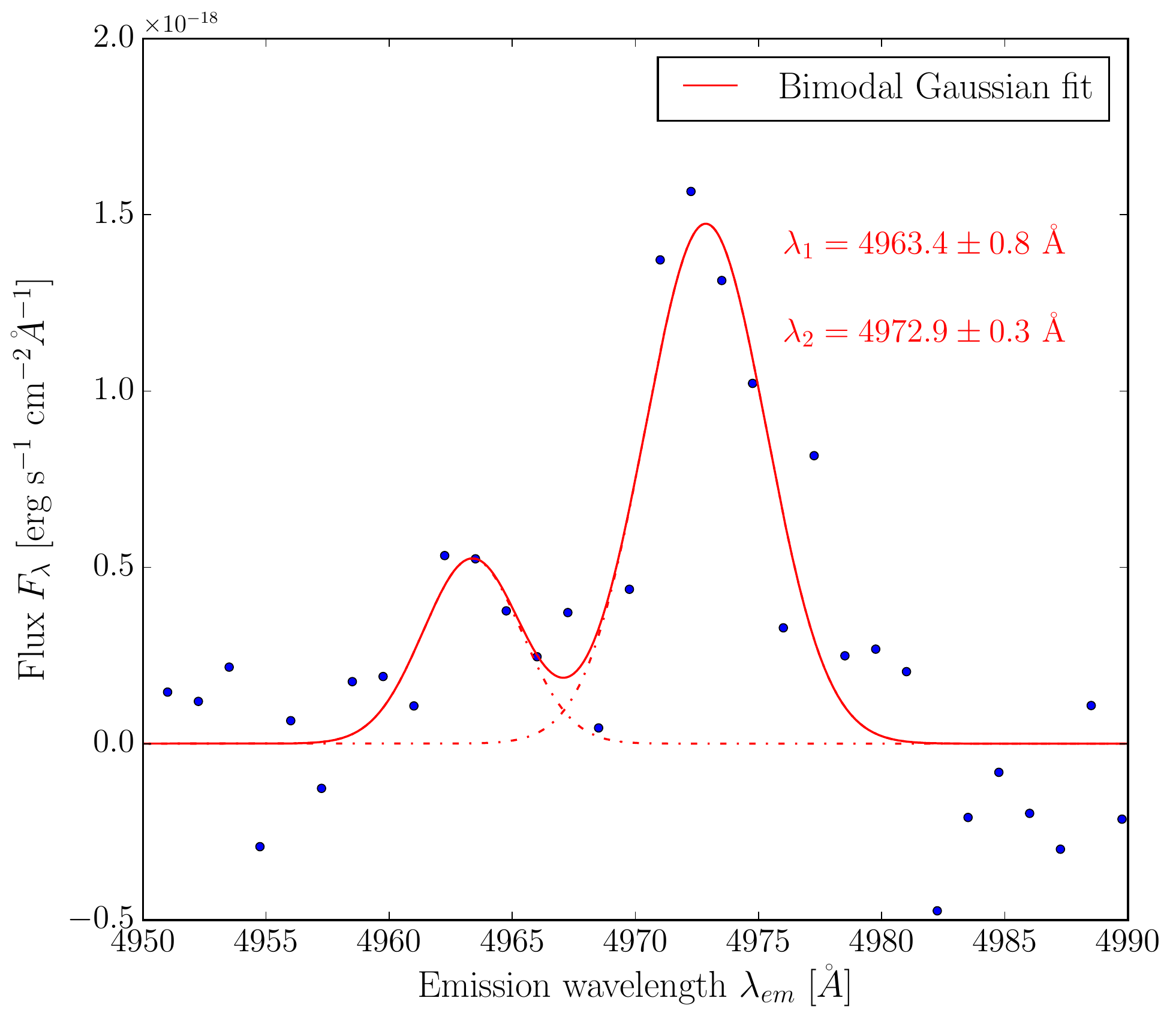}
 \caption{{\sl Left:} Same as Fig.~\ref{fig:LAE1}, but for LAE2.
 {\sl Right:} Same as Fig.~\ref{fig:LAE1}, but for LAE2.
 The spectrum is summed over a disk with $1.0$\arcsec\ radius.}
 \label{fig:LAE2}
\end{figure*}

The second object, LAE2, appears very similar to LAE1 although it is much fainter and
less extended (Fig.~\ref{fig:LAE2}, left panel).
The blue component of the emission line seems brighter compared
to the red component
than it is in LAE1 (see Fig.~\ref{fig:LAE2}, right panel), although it is also more affected by
the noise\footnote{The spectrum is drawn here from a datacube from which continuum sources
were subtracted, thereby avoiding contamination by galaxies close to the line of sight.}.
LAE2 lies $26.5$\arcsec\ away from the QSO, representing a projected distance of
$207$~pkpc and also shares the same redshift as the LAB one. Its coordinates and characteristics
are given in Table~\ref{table:LAE_prop}. 

Its total Ly$\alpha$ flux within a $1.0$\arcsec\ radius is
\begin{equation*}
F_{Ly\alpha} = (12.0 \pm 1.2)\cdot10^{-18} \textnormal{ erg s$^{-1}$ cm$^{-2}$}
\end{equation*}
For the luminosity distance corresponding to the average redshift $z=3.0869$,
we obtain the luminosity
\begin{equation*}
L_{Ly\alpha} = (1.0 \pm 0.1)\cdot10^{42} \textnormal{ erg\,s$^{-1}$}.
\end{equation*}

The rest-frame equivalent width of LAE2 has been estimated in a $r=6$~pixels aperture
at $EW_0\simeq 17\pm 4\,\AA$.

The blue peaks of LAE1 and LAE2 occur at exactly the same redshift, but
this should be considered a mere coincidence, especially because for LAE1, the fit
of the blue peak is very sensitive to noise because the two peaks are not clearly separated.
The presence of these two LAEs in the field probably reflects the overdensity
that is generally related with a QSO.

\subsection{Starburst galaxy near the line of sight of the QSO}
An emission line galaxy is detected about $5.5$\arcsec\ to the NW of the QSO. The
H$_\alpha$, H$_\beta$, [\ion{O}{II}]$\lambda 3727$, [\ion{O}{III}]$\lambda 4959,\,5007$
lines are clearly identified, while the [\ion{N}{II}]$\lambda 6584$ line marginally so.
The five stronger lines allow us to determine
the redshift of this galaxy at $z=0.33005\pm 0.00009$.

We searched the QSO spectrum for a possible sodium absorption linked to this
galaxy in the strong \ion{Na}{I}~D lines at $5889$~\AA\ and $5895$~\AA\ (rest frame). As one
arcsecond corresponds to $4.8$~pkpc at $z=0.33$, the impact parameter is about $26$~pkpc,
and one may expect a non-negligible absorption if the galaxy is not too compact.
We found a faint ($2$\% depth) unresolved line at $7841.8$~\AA\ that might
correspond to \ion{Na}{I}~$\lambda 5889$ (which has the larger oscillator strength),
but it is offset by a velocity $\Delta v=280$\,km\,s$^{-1}$ with respect to the
expected redshifted line. If this feature really were the \ion{Na}{I}~$\lambda 5889$
line, then it would correspond to a \ion{Na}{I} column density on the order of
$10^{11}$~cm$^{-2}$. If all Na were neutral and none of it were locked in dust,
this would correspond to an H column density of about $10^{17}$~cm$^{-2}$ for
solar metallicity. These numbers appear reasonable in view of the Na column
densities ($6\times 10^{11}$~cm$^{-2}$ to $5\times 10^{13}$~cm$^{-2}$) found by
\citet{SM04} for local dwarf starburst galaxies; however, the non-detection of the
\ion{Na}{I}$\lambda 5895$ line makes the above estimates a speculation.
 
\begin{figure}
 \resizebox{\hsize}{!}{\includegraphics{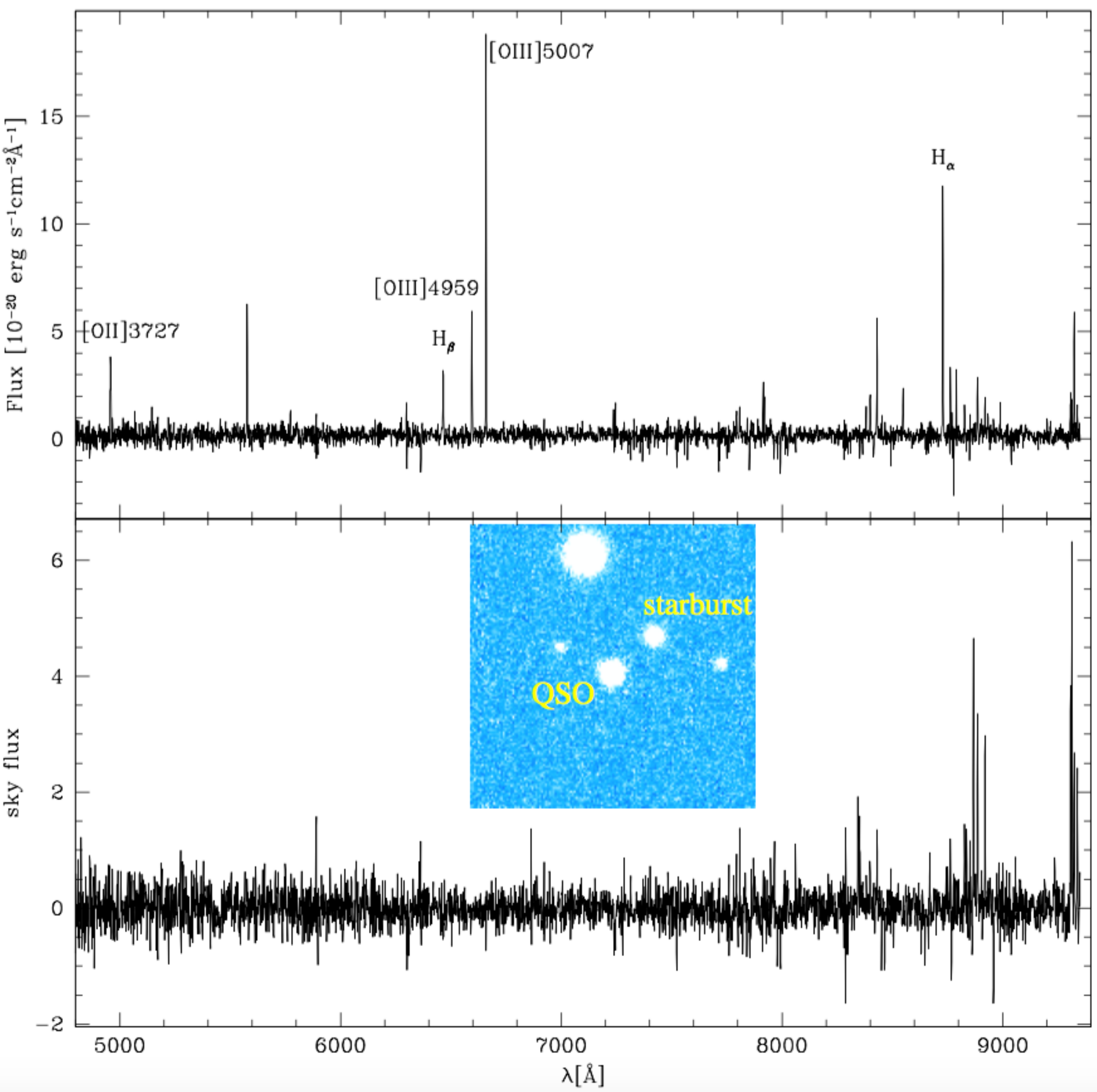}}
 \caption{MUSE spectrum of the starburst galaxy (top) and spectrum of a nearby sky
 background (bottom). The brightest lines are identified. The very bright
 [\ion{O}{i}]\,$\lambda5577.34$ telluric line appears, by chance, in the galaxy spectrum
 but not in the sky spectrum, because it is not subtracted correctly in most parts of
 the datacube. The inset is the monochromatic
 MUSE image at $6658.75$\,\AA\ (redshifted [\ion{O}{iii}] $\lambda 5007$) centered on
 the QSO, showing the position of the emission line galaxy; the image is $30$\arcsec\ across;
 the bright object near the top is a Galactic red giant star.}
 \label{fig:spec2}
\end{figure}

We now focus on other properties of this galaxy. The Balmer decrement can be
estimated through
the ratio H$_\alpha$/H$_\beta \simeq 2.71$: it is slightly lower than the usually
adopted value of $2.86$, suggesting no significant dust reddening. The
[\ion{N}{II}]$\lambda 6584$/H$_\alpha$ and
[\ion{O}{III}]$\lambda 5007$/H$_\beta$ ratios point to a rather typical star-forming
galaxy \citep[see e.g.][Fig. 1]{LMC04}. It has a relatively high excitation:
with a ratio [\ion{O}{III}]$\lambda 5007$/[\ion{O}{II}]$\lambda 3727=3.46$, it
resembles the galaxies of the DR7 sample of \citet{SIM15}, although the
[\ion{O}{III}]$\lambda 4363$ line is not clearly detected.
The [\ion{O}{III}]/[\ion{O}{II}] ratio, although rather high, is lower than that
of the LCEs of \citet{VOS17}.
Its properties are listed in Table~\ref{tab:z0.33} and its spectrum is shown in Fig.~\ref{fig:spec2}.
\begin{table}
\caption{Equatorial coordinates and line intensities of the starburst galaxy at $z=0.33$} 
\label{tab:z0.33}
\begin{center}
\begin{tabular}{lc}
$\alpha_\mathrm{J2000}$ [h:mn:s] & $\delta_\mathrm{J2000}$ [\degr:\arcmin:\arcsec]\\ \hline
$21:02:44.316$                   & $-35:53:02.55$\\ \hline\hline
Line& Observed flux [$10^{-18}$\,erg\,s$^{-1}$\,cm$^{-2}$]\\ \hline
\ion{[O}{II]}$3727$ & $18.1$\\
H$_\beta$                   & $14.9$\\
\ion{[O}{III]}$4959$& $20.1$\\
\ion{[O}{III]}$5007$& $62.6$\\
H$_\alpha$          & $40.3$\\
\ion{[N}{II]}$6584$ & $\sim 2.3$\\ \hline
\end{tabular}
\end{center}
\end{table}

The oxygen abundance can be estimated from the line strengths given in Table~\ref{tab:z0.33}
and from the
O3N2$\equiv\log\left(\frac{\ion{[O}{III]}\lambda5007/H_\beta}{\ion{[N}{II]}\lambda6584/H_\alpha}\right)$
and N2$\equiv\log(\ion{[N}{II]}\lambda6584/H_\alpha)$ criteria \citep[][Eqs. (2) and (4), respectively]{MRS13}.
The main uncertainty, about a factor of two, lies in the \ion{[N}{II]}$\lambda6584$ line intensity.
We obtain
\begin{eqnarray}
12+\log(O/H)&=&8.13\pm 0.07 ~~~~\mathrm{(O3N2)}\\
            &=&8.17\pm 0.14 ~~~~\mathrm{(N2)}
\label{eq:O_abund}
,\end{eqnarray}
assuming a $\pm 0.3$~dex error on both criteria. This corresponds to [O/H]$\sim -0.6$
assuming a solar oxygen abundance of 8.69 \citep{APLA01}, and agrees well with local
starburst galaxies \citep{LS04,FDW16,HSB15,IGF14}.

\section{To which object does the LAB belong?}
Interestingly, the overall spectral shape of the central part of the
LAB displayed in Fig.~\ref{fig:lab_spec} is asymmetric, with a steep blue
side and a less abrupt red side. This typically results from radiative transfer
in the interstellar matter (ISM) of LAE galaxies (see, e.g., Fig.~13 of \citet{KSM11}
and Fig.~5 of \citet{HCB10}). This shape is generally interpreted as the result
of radiative transfer in an expanding medium \citep{VSA08,HVO15,KCC16,YMG17}.

\begin{figure}
 \resizebox{\hsize}{!}{\includegraphics{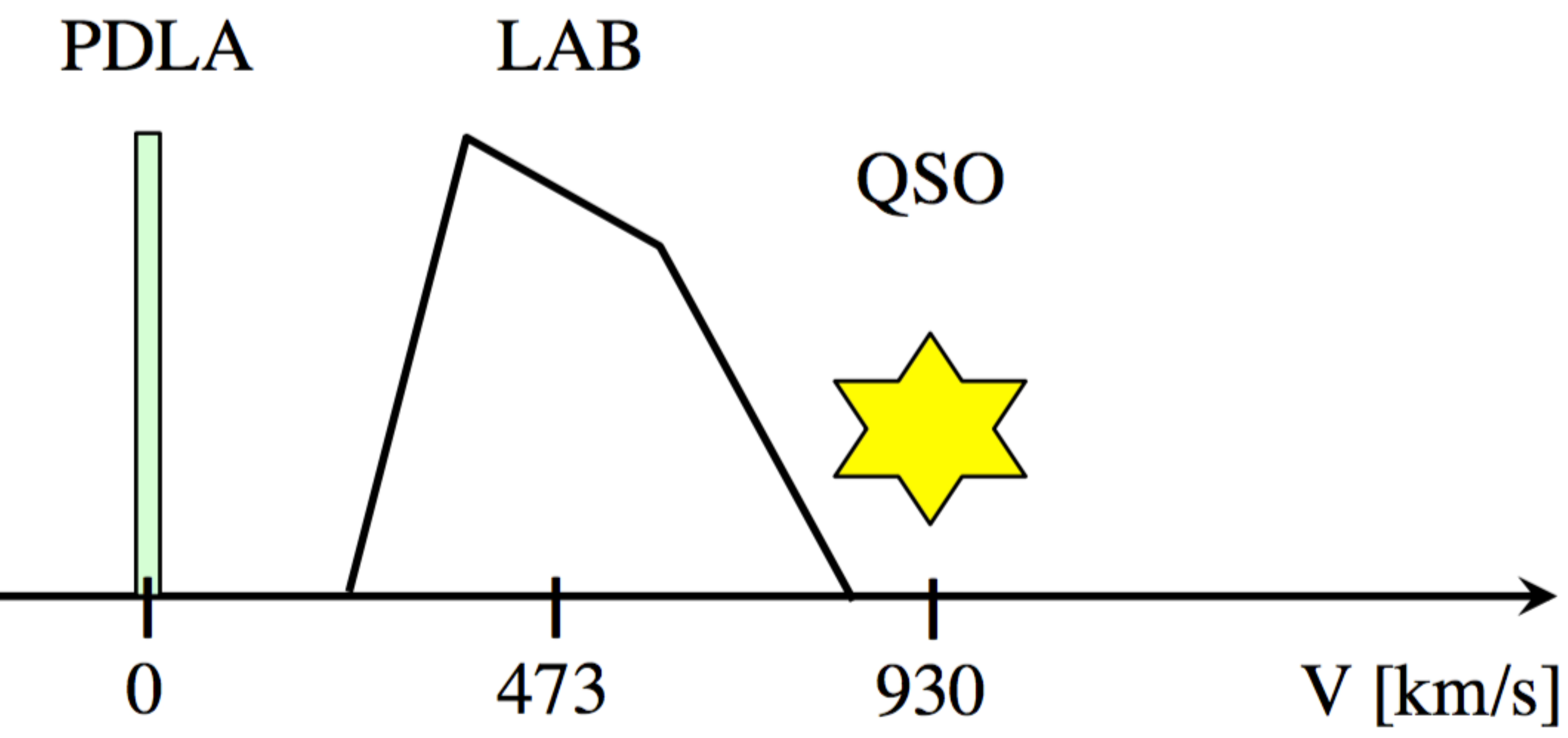}}
 \caption{Schematic view of the positions of the PDLA, LAB, and QSO along
 the velocity axis. The PDLA redshift (determined with metallic
lines)
 was adopted as the origin because of its excellent precision. The QSO lies
 beyond the LAB, while the asymmetry of the LAB emission line would suggest
 the reverse when we assume the model of a QSO that is surrounded by an expanding medium.}
 \label{fig:resp_velocities}
\end{figure}

The crucial point here is that in case of transfer in an expanding medium,
the peak of the Ly$\alpha$ emission is shifted to the red with respect to the
systemic velocity of the emitting galaxy \citep{YMR16,EST14}. Furthermore, the shift is
roughly proportional to the \ion{H}{i} column density \citep[][Fig.~11]{HVO15}.
Now, when we consider the
relative velocities of the PDLA, LAB, and QSO as summarized in Fig.~\ref{fig:resp_velocities},
we see that the average LAB emission line lies redward of the PDLA, but blueward
of the QSO\footnote{An earlier version of this figure has been published
by \citet[][Fig.~6]{LR99}} (even when we admit $z_{QSO}=3.092$, which would place it at
$v=714$\,km\,s$^{-1}$). However, in the case of an expanding medium centered
on the QSO, one would expect the reverse, namely the QSO to lie \textup{{\sl blueward}
}of the LAB emission line. The respective positions of the PDLA and
the LAB are quite consistent with an expanding medium
centered on the PDLA rather than on the QSO. This is true even quantitatively,
since according to the hydrogen column density versus Ly$\alpha$ line redshift relation
mentioned above, we have 
$\Delta v\sim 400-500$\,km\,s$^{-1}$ for the PDLA column density $\log N$(\ion{H}{i})$=20.9$
found in Eq.~\ref{eq:col_dens_zabs}; this is fully consistent with the
$\sim 470$\,km\,s$^{-1}$ shift we see in Fig.~\ref{fig:resp_velocities}.

\begin{figure}
 \resizebox{\hsize}{!}{\includegraphics{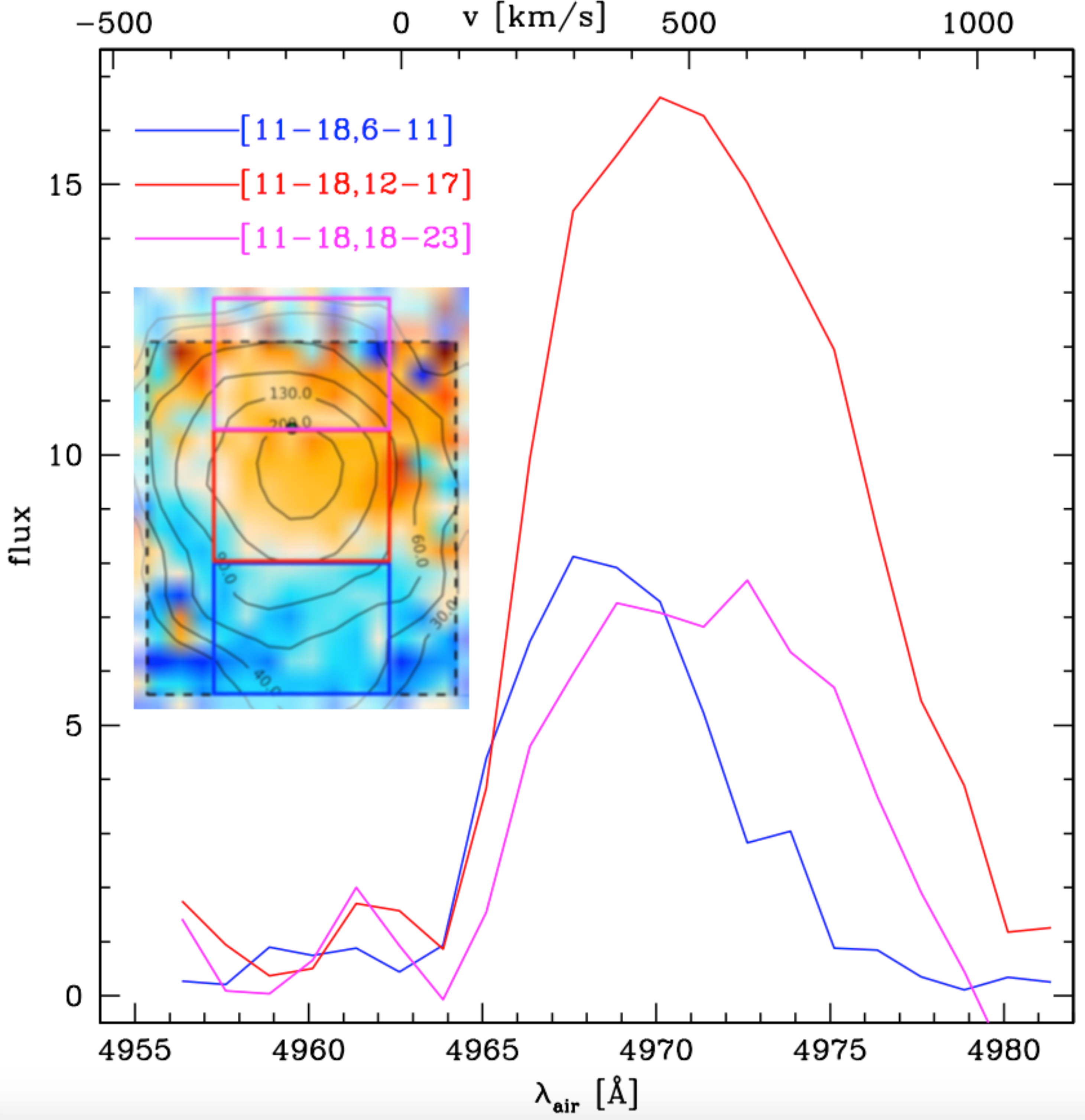}}
 \caption{Ly$\alpha$ line profiles spatially integrated on rectangles covering
 $8\times 6$ spaxels. The coordinates of the lower left and upper right
 corners of the rectangles are given, the [1,1] spaxel being the lower left one
 in each panel of Fig.~\ref{fig:vel_map}; the positions
 of the rectangles are also shown in the inset. The blue, red, and magenta
 curves refer to the southern, middle, and northern rectangle, respectively.}
\label{fig:ly_profile_Vel}
\end{figure}

Thus far, we have neglected the details of the ``velocity'' field displayed in
Fig.~\ref{fig:vel_map} and assumed that the bright central part of the LAB
can be interpreted as an unresolved expanding medium. To explore how far the Ly$\alpha$
line asymmetry depends on position within the LAB, we show in Fig.~\ref{fig:ly_profile_Vel}
the line spatially integrated on three $8\times 6$ spaxel rectangles that are
arranged contiguously from south to north; the
first rectangle covers negative velocities (see Fig.~\ref{fig:vel_map}), while the
other two, lying just to the south and to the north of the QSO, cover
positive velocities. In spite of the overall velocity difference, each Ly$\alpha$
profile displays a similar asymmetry, with a steep blue side and a more gradual
red side. Thus, the skewness of the average line (integrated on the whole LAB)
cannot be explained by the addition of symmetric lines shifted according to the
velocity field, and the expanding medium picture remains essentially valid.

Another completely independent argument in favor of a PDLA-centered LAB stems
from a remark by \citet{HPKZ09} about a similar object, J1240+1455: under the
assumption of an LAB centered on the QSO and lying, therefore, {\sl behind}
the galaxy acting as the PDLA, one would expect to see the PDLA galaxy {\sl in
silhouette} against the bright nebular background. In other words, the PDLA galaxy
would extinct not only the QSO point source, but also part of the LAB, unless
it is so compact as to subtend only a small fraction of an arcsecond, that is, much
less than the seeing disk.
Our observations provide no sign of any extended dark region in the very center of the LAB
that would betray the presence of a large, that is, $\sim 10$\,kpc scale, absorbing
galaxy. Although the average size of late-type galaxies at $z=3$ is in the range
$0.15\arcsec - 0.5$\arcsec\ \citep{FDG04,WFD14}, it is based on the UV continuum;
one can expect that the \ion{H}{i} size is larger, maybe on the order of the seeing,
so that such a galaxy would leave an observable footprint.

Therefore, assuming that the \ion{H}{i} and metal line absorptions do correspond
to the systemic velocity of the PDLA galaxy, there are strong indications that
the LAB surrounds the latter rather than the QSO. This would correspond to the model
(i) envisaged by \citet[][Sect. 4.5]{LR99}. This conclusion is all the more
unexpected, \textcolor[rgb]{1,0.501961,0}{because} each of the $17$ radio-quiet quasars
observed by B16 is surrounded
by an LAB. Why our object should prove an exception remains an open question, although
the presence of the PDLA may betray past interactions that resulted in a ``naked''
QSO like HE0450-2958 \citep{MLC05,EJP09}.

\section{Conclusion}
We have performed an integral-field spectroscopic study of the neighborhood of a radio-quiet QSO
at redshift $z = 3.095$. An emission feature at $z = 3.0887$ is present within the damped Lyman
absorption trough of the QSO spectrum and is spatially located along its line of sight.
According to a previous study \citep{LR99}, this emission feature is probably
a Lyman $\alpha$ blob, although it was not clear to the authors whether it is powered by the QSO
or by the PDLA.
We adapted the method of spectral subtraction of the QSO component, first proposed
by \citet{LR99}, to the MUSE datacube in order to isolate the LAB contribution down
to the position of the QSO itself.
We confirm that the blob presents a sharp velocity gradient with a contrast of
$\sim 200$ km\,s$^{-1}$ close to the line of sight to the QSO (Figs.~\ref{fig:vel_map} and
\ref{fig:vel_map_wide}).

Moreover, we found that the LAB is much more extended
than thought so far, with a filament protruding to the south, so that its total extent
reaches $\sim 120$\,kpc.
The latter result at first sight confirms the finding of B16 that all radio quiet QSOs at $3<z<4$
are surrounded by an LAB, the total extension of which reaches $100$\,kpc or more. The
luminosity of our LAB is also quite typical, even though it lies near the faint end. Furthermore,
like B16, we found no sign of violent kinematics: the FWHM of the Ly$\alpha$ line is
always well below $1000$\,km\,s$^{-1}$, unlike the LABs surrounding high-z radio galaxies.
The upper limits we found on the
intensity of high-ionization lines also compare well with B16. The surface brightness
profile is compatible with a power-law profile, with a slope similar to (but slightly
shallower than) the typical slope found by B16.
However, the asymmetry of the average LAB emission
line suggests that the emitting nebula is centered on the PDLA galaxy rather than
on the QSO because the QSO spectrum lies redward, not blueward of the LAB
emission. Furthermore, we found no sign of any significantly extended dark region in the
very center of the LAB that would betray the presence of the PDLA galaxy in the
foreground, in front of a QSO-centered LAB that would lie in the background.
This makes the comparison with the B16 objects less straightforward
and suggests that our object may be rather uncommon.

Two other LAEs were found close to the NW corner of the MUSE field of view,
at projected distances of $34$\arcsec\ ($265$\,kpc) and $26.5$\arcsec\ ($207$\,kpc)
from the QSO. Although the lack
of other emission lines in the MUSE spectral range leaves some ambiguity, the typical
double profile of the single line, together with a redshift coinciding perfectly
with that of the central LAB, make their Ly$\alpha$ identification most probable. These
other emitters
do not seem to be just other Ly$\alpha$ blobs that would be powered by the QSO only
because the equivalent width of their emission line does not exceed the rest-frame equivalent
width limit (EW$_0 > 240$\,\AA) required to classify them as protogalactic clouds or
dark galaxies \citep{CLH12}

We provided the properties of the PDLA, confirming the \ion{H}{I} column density found
previously by \citet{LR99} with a completely different instrumental setup.

\begin{acknowledgements}
We thank the ESO staff for having made this observation possible. This research has
made use of the SIMBAD database operated at CDS, Strasbourg, France. RAM acknowledges
support by the Swiss National Science Foundation. M.H. acknowledges the support of
the Swedish Research Council (Vetenskapsr{\aa}det) and the Swedish National Space
Board (SNSB), and is a Fellow of the Knut and Alice Wallenberg Foundation.
\end{acknowledgements}

%-------------------------------------------------------------------
\bibliographystyle{aa} % style aa.bst
\bibliography{lab} % your references Yourfile.bib

\end{document}